\documentclass[a4paper,11pt]{article}
\pdfoutput=1 

\usepackage{jheppub} 

\usepackage[utf8]{inputenc}
\usepackage{graphicx}
\usepackage{dcolumn}
\usepackage{bm}

\renewcommand{\figurename}{{\bf Figure}}
\renewcommand{\thefigure}{{\bf\arabic{figure}}}
\renewcommand{\tablename}{{\bf Table}}
\renewcommand{\thetable}{{\bf\arabic{table}}}

\usepackage{natbib}
\usepackage{slashed}
\newcolumntype{x}[1]{>{\centering\arraybackslash\hspace{0pt}}p{#1}}
\usepackage{makecell}
\def\MeV{\,\text{MeV}}
\def\GeV{\,\text{GeV}}
\def\TeV{\,\text{TeV}}

\def\mN{\mathcal{N}}
\def\MPl{M_{\text{Pl}}}
\def\ann{\text{ann}}
\def\bh{\text{bh}}
\def\gc{\text{gc}}
\def\SF{\text{SF}}
\def\ev{\text{ev}}
\def\col{\text{col}}

\title{ \LARGE Gravitational capture of magnetic monopoles by primordial black holes in the early universe}

\author[a]{Chen Zhang}
\author[a,b,c]{and Xin Zhang}

\affiliation[a]{Key Laboratory of Cosmology and Astrophysics (Liaoning) \& College of Sciences, Northeastern University, Shenyang 110819, China}
\affiliation[b]{Key Laboratory of Data Analytics and Optimization for Smart Industry (Ministry of Education), Northeastern University, Shenyang 110819, China}
\affiliation[c]{National Frontiers Science Center for Industrial Intelligence and Systems Optimization, Northeastern University, Shenyang 110819, China}

\emailAdd{zhangchen2@mail.neu.edu.cn}
\emailAdd{zhangxin@mail.neu.edu.cn}

\abstract{It is intriguing to ask whether the existence of primordial black holes (PBHs) in the early universe
could significantly reduce the abundance of certain stable massive particles (SMP) via gravitational capture, after which
the PBHs evaporate before BBN to avoid conflict with stringent bounds. For example, this mechanism is relevant
to an alternative solution of the monopole problem proposed by Stojkovic and Freese, in which magnetic
monopoles produced in the early universe are captured by PBHs, thus freeing inflation from having to occur during
or after the corresponding phase transitions that produced the monopoles. In this work, we reanalyze the solution
by modelling the capture process in the same way as the coexisting monopole annihilation. A subtle issue which is not
handled properly in the previous literature is the choice of an effective capture cross section for diffusive capture. We model this aspect properly and justify our treatment.
A monochromatic PBH mass function and a radiation-dominated era before PBH evaporation are assumed. We find that for Pati-Salam monopoles corresponding to a symmetry breaking scale between
$10^{10}\GeV$ and $10^{15}\GeV$, the capture rate is many orders of magnitude below what is needed to cause a significant
reduction of the monopole density. Within our assumptions, we also find that the magnetic charge that is large enough to make an extremal magnetic black hole cosmologically stable cannot be obtained from magnetic charge fluctuation via monopole capture. The large magnetic charged required by cosmological stability can nevertheless be obtained from magnetic charge fluctuation at PBH formation, and if later the monopole abundance can be reduced significantly by some non-inflationary mechanism, long-lived near-extremal magnetic black holes of observational relevance might result.
}

\begin{document}

\maketitle

\section{Introduction}
\label{sec:Intro}

Stable massive particles (SMPs) whose existence is commonly due to exact or approximate symmetries provide an intriguing
connection between cosmology and particle physics~\cite{Burdin:2014xma}. In the standard model (SM) of particle physics, the approximate baryon number symmetry makes protons cosmologically stable, while the need for baryogenesis in the early universe requires going beyond the SM in a number of directions. Beyond the SM, certain global or gauge symmetries, either discrete or continuous, could give rise to cosmologically stable particles that may act as dark matter (DM), which are crucial for explaining a number of phenomena from galactic to cosmological scales.

The interest of the present work focuses on an interesting class of SMPs---magnetic monopoles~\cite{Dirac:1931kp,tHooft:1974kcl,Polyakov:1974ek}, which arise as a result of a nontrivial second homotopy group $\pi_2 (G/H)$ of the vacuum manifold of some spontaneous symmetry breaking pattern $G/H$ dictated by a (partially) unified gauge theory (see Refs.~\cite{Goddard:1977da,Preskill:1984gd,Vilenkin:2000jqa,Shnir:2005vvi,Weinberg:2012pjx,Mavromatos:2020gwk,ParticleDataGroup:2022pth} for reviews). They can be copiously produced in the corresponding symmetry breaking phase transitions via the Kibble or Kibble-Zurek mechanism~\cite{Kibble:1976sj,Zurek:1985qw}. Being heavy non-relativistic objects, they tend to overclose the universe
during the cosmological evolution if there does not exist an effective mechanism to reduce their number density~\cite{Zeldovich:1978wj,Preskill:1979zi}. Moreover, their relic abundance is more stringently constrained by
the Parker's bound~\cite{Parker:1970xv} coming from the effect of magnetic monopoles on galactic magnetic fields, by the
direct search experiments, and also by catalysis of baryon number violation via the Callan-Rubakov effect~\cite{Rubakov:1982fp,Callan:1982ah,Callan:1982au} depending on
the specific unification models~\footnote{In this work we are concerned with magnetic monopoles associated with the visible electromagnetism. Hidden monopoles associated with some dark gauge symmetry breaking may account for part or all of
dark matter; see e.g. Refs.~\cite{Murayama:2009nj,GomezSanchez:2011orv,Evslin:2012fe,Baek:2013dwa,Khoze:2014woa,Kawasaki:2015lpf,
Nomura:2015xil,Sato:2018nqy,Terning:2019bhg,Daido:2019tbm,Bai:2020ttp,Graesser:2020hiv,Nakagawa:2021nme,Graesser:2021vkr,
Fan:2021ntg,Hiramatsu:2021kvu,Yang:2022quy}.}.

The standard approach to get rid of the overabundance of magnetic monopoles is inflation~\cite{Starobinsky:1980te,Guth:1980zm}, with solving the monopole problem being one of its
most important theoretical motivations. It requires inflation to occur during or after the
corresponding symmetry breaking phase transition (and baryogenesis will be even later), thus establishing a connection
between the particle physics model and the cosmological history. Nevertheless, it is both interesting
and important to ask whether such a connection is inevitable. First, there is the possibility that
gauge coupling unification or even partial unification does not occur, with the side effect that
the elegant explanantion for charge quantization is also lost. Second, in partial unification scenarios
such as the Pati-Salam model~\cite{Pati:1974yy}, it is possible to have a low-scale ($\lesssim 10^{10}\GeV$) strongly first-order Pati-Salam breaking phase transition which could lead to suppressed initial abundance of magnetic monopoles~\cite{Huang:2020bbe}.
Other solutions to the monopole problem include inverse symmetry breaking or symmetry nonrestoration effect in finite-temperature field theory~\cite{Langacker:1980kd,Salomonson:1984rh,Dvali:1995cj,Bajc:1997ky}, eliminating the monopoles
by domain walls which subsequently decay or get destroyed~\cite{Dvali:1997sa,Stojkovic:2005zh}, entropy production effects~\cite{Izawa:1984ww}, and gravitational capture by primordial black holes (PBHs) which evaporate prior to BBN~\cite{Stojkovic:2004hz}. The viability of these alternative solutions may break the connection between inflation and the corresponding symmetry breaking phase transition, and thus allowing for more possibilities of cosmological model
building. However, there are not many alternative solutions and most of them are effective in restrictive portion
of models or parameter space. The viability of these solutions in a more general context, taking into account potential uncertainties in the theoretical modelling and computation, entails further detailed investigations.

In this work we revisit the solution to the monopole problem via gravitational capture by PBHs~\cite{Stojkovic:2004hz}, proposed by Stojkovic and Freese. This idea is of particular interest to us due to two main reasons. First, investigation of monopole capture by PBHs might teach us lessons on whether PBHs may significantly affect the abundance of other SMPs, which may have important implications for physics of dark matter and baryogenesis. Second, recently physics of magnetic black holes has received quite some attention~\cite{Maldacena:2020skw,Bai:2020spd,Liu:2020vsy,Ghosh:2020tdu,Liu:2020bag,Araya:2020tds,Bai:2020ezy,Diamond:2021scl,Chen:2022qvg}. Interesting phenomenological bounds have been obtained~\cite{Ghosh:2020tdu,Bai:2020spd,Diamond:2021scl}, however the formation mechanism of magnetic black holes remains elusive. The difficulty is that one needs to feed a large number of magnetic monopoles into PBHs, while keeping the remaining abundance of monopoles low enough to avoid stringent constraints~\cite{Maldacena:2020skw}. It is conceivable that the mechanism involved in the Stojkovic-Freese (SF) solution might play an important role in some potential formation mechanism of magnetic black holes.

In the SF solution to the monopole problem, two major physical processes that affect the monopole abundance are monopole annihilation and gravitational capture by PBHs. These two processes are in fact quite similar. Both processes are driven by long-range forces obeying an inverse-square law, and in both processes the movement of monopoles in the primordial plasma is that of a Brownian motion. In the SF paper~\cite{Stojkovic:2004hz}, however, the gravitational capture by PBHs is modelled somewhat differently from the monopole annihilation. In this work, we have instead modelled the gravitational capture by PBHs in the same manner as monopole annihilation. We find that our modelling leads to significantly smaller capture rates compared to the SF modelling. The difference can be traced to the fact that the SF modelling uses an effective capture cross section that is
not appropriate in the diffusive regime. Moreover, we recognize that assuming radiation domination, the use of an appropriately extended PBH mass function in the spirit of that used in Ref.~\cite{Stojkovic:2004hz} instead of a monochromatic one employed in our study should improve significantly the efficiency in reducing the monopole abundance.

It is possible for PBHs to acquire some magnetic charge at formation if the formation temperature is below the symmetry breaking phase transition temperature. Even if the PBHs do not carry magnetic charge at formation, the monopole capture process may leave a residual magnetic charge on the PBH because there is fluctuation on the number of absorbed monopoles and antimonopoles. This magnetic charge fluctuation is expected to leave a magnetic charge of about $\sqrt{N}$ if the total number of absorbed monopoles and antimonopoles is $N$ for each PBH. The fate of this residual magnetic charge (including any initial magnetic charge) depends on its magnitude. For a sufficiently large residual magnetic charge ($\gtrsim\mathcal{O}(10^6))$ for a monopole mass ($\sim 10^{17}\GeV$), the PBH evolves toward a magnetically charged extremal Reissner-Nordström (RN) black hole that is cosmologically stable~\cite{Maldacena:2020skw}. Otherwise, its evaporation should be qualitatively similar to that of an uncharged PBH with the same mass, but at the final stage it is unstable against decaying into multiple magnetic monopoles which again should be taken into account when we compute the final monopole abundance. In this work we also examine the magnitude of residual magnetic charge. Assuming a monochromatic PBH mass function and a radiation-dominated universe before PBH evaporation, we find that for the monopole capture processes, it is not possible to get a sufficiently large residual magnetic charge required by cosmological stability. We show that the large magnetic charge required by cosmological stability might be obtained already at PBH formation\footnote{In such a case we only require radiation domination before PBH formation rather than PBH evaporation.}. Although with the monochromatic PBH mass function the monopole problem is not solved yet, such an investigation may lead to insights about the formation of magnetic black holes in the early universe.

It is interesting to note that evaporating PBHs create hot spots which could reproduce magnetic monopoles during their cooling~\cite{Das:2021wei,He:2022wwy}, an effect that is unknown at the time of the SF proposal. According to the latest study~\cite{He:2022wwy}, the highest temperature achieved in the hot spot is not larger than $\mathcal{O}(10^{10}\GeV)$,
assuming a fine structure constant of about $0.1$. Therefore, for simplicity, in this work we consider the symmetry breaking phase transition scale to be $\gtrsim\mathcal{O}(10^{10}\GeV)$.

This work is organized as follows. In Sec.~\ref{sec:rev} we briefly review the production of magnetic monopoles in a symmetry breaking phase transition in the early universe, and the subsequent monopole-antimonopole annihilation. In Sec.~\ref{sec:gc} we present our modelling of the gravitational capture of monopoles by PBHs. The essential difference compared to the SF modelling consists in the choice of effective capture cross section in the diffusive capture regime, which we explain in detail in Sec.~\ref{subsec:flux}. In Sec.~\ref{sec:mcf} we outline the computation of magnetic charge fluctuation. In Sec.~\ref{sec:prep} we collect the ingredients required for comparing the different ways of modelling and analyzing the consequences of magnetic charge fluctuations which are performed in Sec.~\ref{sec:abp}. We present the discussion and conclusions in Sec.~\ref{sec:dnc}.

\section{Review of monopole production and annihilation}
\label{sec:rev}

\subsection{Monopole production}

A magnetic monopole can be viewed as an extended field configuration stabilized by nontrivial topology associated with the mapping from spatial infinity to the vacuum manifold of some spontaneous symmetry breaking. The order parameter of the symmetry breaking phase transition lives on the vacuum manifold and it relies on local interactions to align order parameters of nearby regions. In a cosmological setting, the range of interactions is limited by the particle horizon, and thus different horizon patches will choose their order parameter values independently. At the junctions of multiple horizon patches there is some probability to form field configurations with a nontrivial winding number, resulting in the production of magnetic monopoles. This is the basic picture of producing topological defects in the early universe via the Kibble mechanism~\cite{Kibble:1976sj}. Note that in gauge theories it is possible to formulate the above discussion in a gauge-invariant manner~\cite{Weinberg:2012pjx}.

For definiteness, we consider a radiation-dominated universe. The energy density $\rho$ and entropy density $s$ at temperature $T$ are given by
\begin{align}
\rho=K_1 T^4,\quad s=K_2 T^3,
\label{eqn:rhos}
\end{align}
with
\begin{align}
K_1=\frac{\pi^2}{30}\mN,\quad K_2=\frac{2\pi^2}{45}\mN,
\label{eqn:k1k2}
\end{align}
with $\mN$ being the number of effective relativistic degrees of freedom at temperature $T$. Here for simplicity we
approximate $\mN$ appearing in $\rho$ and $s$ as the same, and neglect the change of $\mN$ with temperature. $\mN$ typically ranges from  $100$ to $1000$ for temperatures above the electroweak scale, depending on the specific particle physics model.
According to the Friedmann equation, the Hubble parameter can be expressed as
\begin{align}
H=K\frac{T^2}{\MPl},
\label{eqn:hubble}
\end{align}
with
\begin{align}
K=\bigg(\frac{4\pi^3\mN}{45}\bigg)^{1/2},
\label{eqn:k}
\end{align}
and the Planck mass ($G$ is the Newton constant)
\begin{align}
\MPl\equiv G^{-1/2}=1.2\times 10^{19}\GeV=2.2\times 10^{-5} \text{g}.
\label{eqn:Planckmass}
\end{align}
The particle horizon $d_H$ is given by the inverse of the Hubble parameter
\begin{align}
d_H=H^{-1}.
\end{align}
On average each volume of $d_H^3$ should contain $p_M$ monopoles, with $p_M$ being a number that is not much less than $1$.
Therefore the monopole number density $n_M$ at production should satisfy
\begin{align}
n_M(T_c)\gtrsim p_M d_H^{-3}(T_c)=p_M H^3(T_c)=p_M K^3\frac{T_c^6}{\MPl^3}.
\end{align}
Here $T_c$ denotes the temperature at which the monopoles are produced. As an approximation in this work we identify
it with the critical temperature of the phase transition, although strictly speaking it can be somewhat lower than the
true critical temperature~\cite{Weinberg:2012pjx}. We now define the monopole yield $r$ as
\begin{align}
r\equiv\frac{n_M}{s},
\end{align}
then the monopole yield at $T=T_c$, denoted $r_i$ hereafter, satisfies
\begin{align}
r_i\equiv r(T_c)=\frac{n_M(T_c)}{s(T_c)}\gtrsim p_M K^3 K_2^{-1}\bigg(\frac{T}{\MPl}\bigg)^3.
\end{align}
This just gives the Kibble estimate, which can be expressed as
\begin{align}
r_i\gtrsim p (8\pi)^{3/2}\mN^{1/2}\bigg(\frac{T_c}{\MPl}\bigg)^3,
\label{eqn:kibble}
\end{align}
where
\begin{align}
p\equiv p_M\frac{\pi}{12\sqrt{10}}
\label{eqn:defp}
\end{align}
is a number not much less than $0.1$.

It should be emphasized that the Kibble estimate only gives a lower bound on the initial monopole abundance,
while the actual initial abundance can be much larger. There are three scenarios that can be envisioned which we discuss below: first-order phase transitions, second-order phase transitions, and crossover phase transitions.

\vspace{5mm}
\textbf{(i) First-order phase transitions}

In the case of first-order phase transitions, the phase transition proceeds by bubble nucleation. Inside a single bubble
the order parameter should be uniform, while order parameters in different bubbles should be uncorrelated. Suppose the characteristic bubble size at bubble coalescence is $R$, then the number density of monopoles at production is estimated to be
\begin{align}
n_M\simeq p_M R^{-3}.
\end{align}
$R$ is related to the parameter $\beta$ that characterizes the inverse duration of the phase transition (see Ref.~\cite{Hindmarsh:2020hop} for definition and discussion) and the bubble wall velocity $v_w$ as
\begin{align}
R=\frac{(8\pi)^{1/3}v_w}{\beta}.
\end{align}
Introducing the dimensionless version of the $\beta$ parameter
\begin{align}
\tilde\beta\equiv\frac{\beta}{H(T_p)},
\end{align}
with $T_p$ being the percolation temperature, it is possible to express $r_i$ as
\begin{align}
r_i\simeq p(\tilde\beta v_w^{-1})^3 (8\pi)^{3/2}\mN^{1/2}\bigg(\frac{T_c}{\MPl}\bigg)^3.
\label{eqn:ri1st}
\end{align}
Again $p$ is some number not much less than $0.1$ and for our purpose we have made the
approximation $T_p\approx T_c$. For a strongly first-order phase transition, one gets
$\tilde\beta v_w^{-1}\simeq\mathcal{O}(1)$, so that Eq.~\eqref{eqn:ri1st} just saturates
the Kibble estimate in Eq.~\eqref{eqn:kibble}. For a weakly first-order phase transition,
with typical values $\tilde\beta v_w^{-1}\simeq\mathcal{O}(10\sim10^3)$, obviously the monopole yield can
be orders of magnitude larger than the Kibble estimate.

\vspace{5mm}
\textbf{(ii) Second-order phase transitions}

In the case of second-order phase transitions, monopole density is determined via the Kibble-Zurek mechanism (see Refs.~\cite{Vachaspati:2006zz,delCampo:2013nla,Graesser:2020hiv} for reviews). The basic picture is as follows.
What replaces $R$ in the case of first-order phase transitions should be some correlation length. Usually, the
correlation length diverges at the critical point. However the cosmic expansion in the early universe determines
a finite quench time $\tau_Q$ which is just the inverse of the Hubble parameter. To discuss the relevant physics,
we introduce the reduced distance parameter $\epsilon$, defined as
\begin{align}
\epsilon=\frac{\omega_c-\omega}{\omega_c},
\end{align}
with $\omega$ being some control parameter (such as temperature), and $\omega_c$ being its value at the critical
point. Obviously, $\epsilon$ characterizes how close the system is to the critical point. As $\epsilon\rightarrow 0$, the \emph{equilibrium} correlation
length $\xi$ and the \emph{equilibrium} relaxation time $\tau$ then scale as
\begin{align}
\xi(\epsilon)=\frac{\xi_0}{|\epsilon|^\nu},\quad\tau(\epsilon)=\frac{\tau_0}{|\epsilon|^{\mu}},
\label{eqn:nudef}
\end{align}
with $\mu,\nu$ being two critical exponents. We assume a linear quench, that is (we take $t=0$ corresponding to the critical temperature $T=T_c$)
\begin{align}
\epsilon(t)=\frac{t}{\tau_Q},\quad\text{for}\,t\in [-\tau_Q,\tau_Q].
\end{align}
Now the crucial thing is to note that there is a specific time $\hat t$, defined in such a way that
\begin{align}
\tau(\hat t)=\hat t,
\end{align}
which means that the equilibrium relaxation time is about the same as with the time elapsed after crossing
the critical point. $\hat t$ is known as the freeze-out time, since in the Kibble-Zurek mechanism
it is assumed that during the time interval $[-t,t]$ the dynamics (i.e. the correlation length $\xi$)
is frozen while outside the $[-t,t]$ interval, the dynamics follows the usual adiabatic behavior. Thus,
the maximum correlation length reached during the second-order phase transition is
\begin{align}
\hat\xi\equiv\xi(\hat\epsilon),\quad\hat\epsilon\equiv|\epsilon(\hat t)|.
\end{align}
It is then straightforward to obtain
\begin{align}
\hat\xi=\xi_0\bigg(\frac{\tau_Q}{\tau_0}\bigg)^\frac{\nu}{1+\mu},\quad\tau_Q=\frac{1}{H(T_c)}.
\label{eqn:hatxi}
\end{align}
Each volume of $\hat\xi^3$ should contain about $1$ monopole, thus the estimated initial monopole number density
is
\begin{align}
n_M(T_c)\simeq\frac{1}{\xi_0^3}\bigg(\frac{\tau_0}{\tau_Q}\bigg)^\frac{3\nu}{1+\mu}.
\label{eqn:nm2nda}
\end{align}
To proceed further, suppose the second-order phase transition is driven by some scalar field dynamics
with a dimensionless quartic coupling $\lambda$ and the thermal effective mass $m_\sigma$ for a typical
scalar resonance. Then we may approximate
\begin{align}
\xi_0\simeq\tau_0\simeq m_\sigma^{-1}\simeq\frac{1}{\sqrt{\lambda}T_c},
\end{align}
then Eqs.~\eqref{eqn:hatxi}--\eqref{eqn:nm2nda} can be expressed as
\begin{align}
\hat\xi &\simeq\frac{1}{\sqrt{\lambda}T_c}\bigg[\frac{\sqrt{\lambda}T_c}{H(T_c)}\bigg]^{\frac{\nu}{1+\mu}}, \\
n_M(T_c) &\simeq\big(\sqrt{\lambda}T_c\big)^3\bigg[\frac{H(T_c)}{\sqrt{\lambda}T_c}\bigg]^{\frac{3\nu}{1+\mu}}.
\end{align}
Using Eqs.~\eqref{eqn:rhos}--\eqref{eqn:k} we may obtain
\begin{align}
r_i\simeq\lambda^{3/2}K_2^{-1}\bigg[\frac{KT_c}{\sqrt{\lambda}\MPl}\bigg]^{\frac{3\nu}{1+\mu}}.
\label{eqn:ri2ndb}
\end{align}
We will typically consider
\begin{align}
\mu=\nu,\quad \lambda\simeq 1,
\end{align}
then
\begin{align}
r_i\simeq 0.02\bigg(\frac{17T_c}{\MPl}\bigg)^\frac{3\nu}{1+\nu},\quad\text{for}\,\,\mN=100,~\lambda=1.
\label{eqn:KZ}
\end{align}
Typical value of the critical exponent $\nu$ is $0.5\sim 0.8$~\cite{Murayama:2009nj}, leading to
an initial monopole abundance orders of magnitude larger than the bound set by the Kibble estimate \eqref{eqn:kibble}.

\vspace{5mm}
\textbf{(iii) Crossover phase transitions}

One can model the dynamics of a crossover phase transition in the same spirit as that
of a second-order phase transition, with appropriate modification to the scaling behavior
of $\xi$ and $\tau$ so that their critical scaling behavior tapers off very close to $\epsilon=0$~\cite{Zurek:1999ym}.
In such a case it is possible that the correlation length saturates at an even smaller value
compared to the case of a second-order phase transition, leading to a larger rate of defect formation.
Lacking a generally accepted way to model the crossover phase transition, here we will not attempt
to estimate the initial monopole abundance in crossover phase transitions in a more quantitative manner.

\subsection{Monopole annihilation}

Consider a magnetic monopole with mass $m$ that carries a magnetic charge of
\begin{align}
\chi g,
\label{eqn:chidef}
\end{align}
where $\chi$ is an integer
(for a unit magnetic monopole, $\chi$=1), and $g$ is the unit magnetic charge in the natural Gaussian system, that is
\begin{align}
g=\frac{1}{2e},\quad e=\sqrt{\alpha}.
\label{eqn:unitmag}
\end{align}
Using $\alpha=\frac{1}{137}$ we obtain that
\begin{align}
g\simeq 5.9.
\end{align}
When a magnetic monopole moves with velocity $\textbf{v}$ the primordial plasma in the early universe, it experiences a drag force which can be expressed as~\cite{Vilenkin:2000jqa}
\begin{align}
\textbf{F}_\text{drag}=-CT^2f(v)\textbf{v},
\label{eqn:dragforce}
\end{align}
with $f(v)$ being a slowly varying function with $f(0)=1$ and $f(1)=3/2$, and
\begin{align}
C=\frac{2\pi}{9}\bar{C}\chi^2 g^2\sum_a b_a e_a^2,
\label{eqn:Cdef}
\end{align}
where the summation is over the spin states of light charged particles, $b=1$ for bosons and $b=\frac{1}{2}$ for fermions,
and $\bar C$ is an angular integral that is roughly $5\sim 10$. It is expected that $C\chi^{-2}\sim(1-5)\mN_c$, with
$\mN_c$ being the number of relativistic effective charged degrees of freedom~\cite{Vilenkin:2000jqa}.

In the following we approximate $f(v)=1$, then the equation of motion of a nonrelativistic magnetic monopole in the presence of the drag force is
\begin{align}
m\dot{\textbf{v}}=-CT^2\textbf{v}.
\end{align}
Its solution $ \textbf{v}= \textbf{v}_0e^{-t/\tau_M}$ is characterized by a timescale $\tau_M$
\begin{align}
\tau_M=\frac{m}{CT^2}.
\label{eqn:taum}
\end{align}
$\tau_M$ can be regarded as the characteristic timescale the monopole needs to forget its initial velocity,
or the mean free time of the monopole. We may then derive the monopole mean free path $\ell$
by multiplying it with the monopole's thermal velocity $v_T\sim(3T/m)^{1/2}$\footnote{Magnetic monopoles are not in chemical equilibrium with the plasma, but due to the electromagnetic interactions with the charged particles,
they are kept in kinetic equilibrium, leading to a thermal velocity of $\sim(3T/m)^{1/2}$.}. The result is
\begin{align}
\ell\simeq\frac{1}{CT}\bigg(\frac{m}{T}\bigg)^{1/2}.
\end{align}
One should note however the mean free time $\tau_M$ and the mean free path $\ell$ do not correspond to
the average time spent and distance traversed by the monopole during two scatterings with the plasma particles.
Instead, from the derivation above we see that they correspond to the average time spent and distance traversed by
the monopole to have a significant change of its velocity.

Now besides the drag force exerted by the plasma, we consider the attractive force between a monopole and an antimonopole
at a distance $\bar{R}$. When the attractive force is balanced by the drag force, the monopole will attain a drift velocity
\begin{align}
v_\text{D}\simeq\frac{\chi^2 g^2}{CT^2 \bar{R}^2}.
\end{align}
For the typical monopole separation $d_\text{ann}\sim n_M^{-1/3}$, we may estimate the capture time as
\begin{align}
\tau_\text{ann}\simeq \frac{d_\text{ann}}{v_\text{D}(d_\text{ann})}=\frac{CT^2}{\chi^2 g^2 n_M}.
\end{align}
This implies that we may express the time evolution of the monopole number density $n_M$ as
\begin{align}
\dot{n}_M=-Dn_M^2-3\frac{\dot{a}}{a}n_M,
\label{eqn:nmevolve}
\end{align}
with
\begin{align}
D=\frac{1}{\tau_\text{ann} n_M}=\frac{\chi^2 g^2}{CT^2},
\label{eqn:D}
\end{align}
and the $-3\frac{\dot{a}}{a}n_M$ term obviously takes into account of the effect of cosmic expansion.
Eq.~\eqref{eqn:nmevolve} can be solved by considering the evolution of $r=n_M/s$ with respect to
the temperature $T$. Introducing the reduced temperature variable $z$,
\begin{align}
z(T)\equiv\frac{T}{T_c},
\end{align}
the evolution of $r$ from $z_1\equiv(T_1)$ to $z_2\equiv(T_2)$ can be expressed as
\begin{align}
r_2=\bigg[\frac{1}{r_1}+\frac{\Delta}{T_c}\bigg(\frac{1}{z_2}-\frac{1}{z_1}\bigg) \bigg]^{-1},
\label{eqn:r1r2ann}
\end{align}
where $r_1=r(z_1),r_2=r(z_2)$, and $\Delta$ is given by
\begin{align}
\Delta\equiv K_2 K^{-1}\MPl DT^2=K_2 K^{-1}C^{-1}\chi^2 g^2\MPl.
\label{eqn:delta}
\end{align}
Note that $\Delta$ is a temperature-independent quantity with mass dimension $1$.

Since $D\propto T^{-2}$ (see Eq.~\eqref{eqn:D}), the diffusive capture is more efficient at low
temperature than at high temperature. This can be traced to the fact that at low temperature,
the smaller drag force allowed a larger drift velocity of the monopole, thus reducing the capture time.
Once the monopole and the antimonopole are captured into a Coulomb bound state, then they cannot
escape annihilation in the diffusive environment.

However, the above picture only holds when the monopole mean free path is smaller than the capture radius.
The capture radius $r_c^{\text{ann}}$ is found by letting the negative Coulomb potential energy be comparable
to the monopole's thermal kinetic energy. Therefore
\begin{align}
r_c^{\text{ann}}\simeq\frac{\chi^2 g^2}{T}.
\end{align}
Requiring $\ell<r_c^{\text{ann}}$ leads to
\begin{align}
T>T_\text{ann},\quad T_\ann=\frac{m}{C^2\chi^4 g^4}.
\label{eqn:Tann}
\end{align}
Thus, if the temperature drops below $T_\ann$, diffusive capture of monopoles will cease to be effective, and
monopoles and antimonopoles can only annihilate by radiative capture via bremsstrahlung emission, leading to
a much smaller capture rate which we neglect in this study~\cite{Weinberg:2012pjx}.

We introduce the reduced variables
\begin{align}
\delta=\frac{m}{T_c},\quad x=\frac{T_c}{\MPl},
\label{eqn:deltax}
\end{align}
then using Eq.~\eqref{eqn:r1r2ann} for an initial monopole yield $r_i$ at $T=T_c$, the final monopole yield $r_\ann$ at $T=T_\ann$ is approximately given by
\begin{align}
r_\ann\simeq\min(r_i,r_\star),\quad r_\star\equiv K_2^{-1}KC^{-1}\chi^{-6}g^{-6}\delta x.
\end{align}
This means that if $r_i<r_\star$, then the monopole yield cannot be reduced further through annihilation.
On the other hand, if $r_i>r_\star$, annihilation can always reduce the monopole yield to $r_\star$ which
is independent of the initial yield, and the annihilation is most efficient at temperature close to $T_\ann$.
Note also that $r_\star$ is very sensitive to $\chi$ and $g$.

\section{Capture of monopoles by primordial black holes}
\label{sec:gc}

\subsection{The SF modelling of monopole capture by PBHs}

First consider PBHs with a universal mass $m_\bh$ (so the Schwarzchild radius is $R_\bh=2m_\bh\MPl^{-2}$),
uniformly distributed in the early universe with number density $n_\bh$. In the SF paper~\cite{Stojkovic:2004hz}, the gravitational capture of monopoles by PBHs is modelled by a new term in the evolution equation for $n_M$, so that
Eq.~\eqref{eqn:nmevolve} is modified to be
\begin{align}
\dot{n}_M=-Dn_M^2-F_\text{SF}n_M-3\frac{\dot{a}}{a}n_M,
\end{align}
where
\begin{align}
F_\text{SF}\equiv n_\bh\sigma_g v_M,
\label{eqn:fsfdef}
\end{align}
with $v_M$ being the average incident velocity
of monopoles on a PBH, and $\sigma_g$ being the gravitational capture cross section, given by~\cite{Frolov:2011bhp}
\begin{align}
\sigma_g=4\pi v_M^{-2}R_\bh^2,\quad R_\bh=2m_\bh\MPl^{-2}.
\label{eqn:sgvm}
\end{align}

Then the key issue is to determine $v_M$. One naturally tries to link $v_M$ with
the monopole thermal velocity $v_T=(3T/m)^{1/2}$. In the SF modelling, $v_M$ is taken to
be a random walk velocity, such that if there is just one random walk step, $v_M$ is equal to
$v_T$. $v_M$ and $v_T$ are then related by
\begin{align}
v_M=\frac{v_T}{\sqrt{N}},
\label{eqn:vmvt}
\end{align}
in which $N$ is the number of random walk steps. $N$ is determined as follows. For the gravitational
capture of monopoles by PBHs, one may determine a capture radius $r_c^\gc$ in a similar manner
as $r_c^\ann$. That is, the negative gravitational energy should be comparable to the monopole's
thermal kinetic energy when the distance between the monopole and the PBH is $r_c^\gc$,
\begin{align}
r_c^\gc=\frac{mm_\bh}{\MPl^2 T},
\label{eqn:rcgc}
\end{align}
$\sqrt{N}$ is then taken as the ratio between $r_c^\gc$ and the monopole mean free path $\ell$,
\begin{align}
\sqrt{N}=\frac{r_c^\gc}{\ell}.
\label{eqn:sqrtn}
\end{align}
Thus, $v_M$ can be viewed as the random walk velocity for the monopole to generate a root-mean-square
displacement of about the capture radius $r_c^\gc$. Using $v_T=(3T/m)^{1/2}$, and Eqs.~\eqref{eqn:vmvt} and \eqref{eqn:sqrtn},
we obtain the incident velocity $v_M$ as
\begin{align}
v_M\simeq\frac{\sqrt{3}\MPl^2}{Cmm_\bh}.
\label{eqn:incidentSF}
\end{align}
Interestingly, the dependence on $T$ disappears in the expression for $v_M$. The gravitational capture
cross section then becomes
\begin{align}
\sigma_g\simeq \frac{16}{3}\pi C^2\frac{m^2 m_\bh^4}{\MPl^8}.
\label{eqn:sigmag}
\end{align}
By introducing two parameters $f_\SF,\beta_\SF$, defined through
\begin{align}
n_\bh=\frac{f_\SF\mN T^4}{m_\bh},\quad m_\bh=\beta_\SF\times 0.2\mN^{-1/2}\frac{\MPl^3}{T^2},
\end{align}
one may express $F_\SF$ in Eq.~\eqref{eqn:fsfdef} as
\begin{align}
F_\SF\simeq f_\SF\beta_\SF^2 Cm.
\label{eqn:FSF1}
\end{align}
Note that $f_\SF$ characterizes the energy density fraction of PBHs, while $\beta_\SF$ characterizes
the ratio between a single PBH mass and the mass within the particle horizon. Assuming a monochromatic
PBH mass function and that the accretion does not change the PBH mass significantly, we expect
\begin{align}
f_\SF\propto T^{-1},\quad \beta_{\SF}\propto T^2,\quad \text{monochromatic},
\end{align}
so that $f_\SF\beta_\SF^2\propto T^3$. Nevertheless in the SF paper~\cite{Stojkovic:2004hz} the assumption
on the PBH mass function is such that
\begin{align}
f_\SF=\text{const},\quad \beta_{\SF}=\text{const},\quad \text{SF mass function}.
\label{eqn:sfassumption}
\end{align}
In order for Eq.~\eqref{eqn:sfassumption} to hold, new PBHs must keep forming while the old PBHs
must be keep evaporating in some manner to keep $f_\SF,\beta_\SF$ constant. As was pointed
out in Ref.~\cite{Stojkovic:2004hz}, $f_\SF$ and $\beta_\SF$ then represent some average feature of the
system.

For definiteness we only use the phrase ``SF modelling'' to refer to the treatment of the gravitational capture
of monopoles by PBHs in the SF paper~\cite{Stojkovic:2004hz} without any assumption on the PBH mass function.
The specific requirement in Eq.~\eqref{eqn:sfassumption} will be termed as the ``SF mass function''.


In retrospect, the use of Eq.~\eqref{eqn:sgvm} as the effective capture cross section in the diffusive regime
is subject to one serious criticism. Eq.~\eqref{eqn:sgvm} is derived from
solving the geodesic evolution of a massive particle in the Schwarzschild geometry, without the frequent bombardment
from other particles at all~\cite{Frolov:2011bhp}. In the diffusive regime the monopole is likely to be deflected
through collisions with other particles along its journey to the PBH event horizon, even it is already inside the capture
radius. Therefore we set out to find alternative modelling of the gravitational capture process.

\subsection{The drift modelling of monopole capture by PBHs}

As noted previously, the gravitational capture of monopoles by PBHs are quite similar to the monopole-antimonopole annihilation. We are thus motivated to model the two processes in a similar manner. Therefore, let us consider the
balancing between the gravitational force and the drag force exerted on a monopole, which determines a monopole drift
velocity $u_\text{D}$ as a function of the monopole-PBH distance $\bar{R}$
\begin{align}
u_\text{D}(\bar{R})=\frac{m_\bh m}{\MPl^2}\frac{1}{CT^2\bar{R}^2}.
\end{align}
The typical separation between PBHs is $n_\bh^{-1/3}$, thus we will use a typical drift velocity
for the capture process as
\begin{align}
u_\text{D}(n_\bh^{-1/3})=\frac{m_\bh m}{\MPl^2}\frac{n_\bh^{2/3}}{CT^2}.
\end{align}
Thus the typical capture time is
\begin{align}
\tau_\gc=\frac{n_\bh^{-1/3}}{u_\text{D}(n_\bh^{-1/3})}=\frac{\MPl^2 CT^2}{n_\bh m_\bh m}.
\label{eqn:captime}
\end{align}
The typical capture frequency per monopole is
\begin{align}
F\equiv\tau_\gc^{-1}=\frac{n_\bh m_\bh m}{\MPl^2 CT^2}.
\label{eqn:fe1}
\end{align}
The evolution of the monopole number density satisfies
\begin{align}
\dot{n}_M=-Dn_M^2-F n_M-3\frac{\dot{a}}{a}n_M.
\label{eqn:nmdrift}
\end{align}

In analogy with monopole annihilation, gravitational capture of monopoles by PBHs
should be effective only when the monopole mean free path $\ell$ is less than the gravitational
capture radius
\begin{align}
r_c^\gc=\frac{m_\bh m}{\MPl^2 T}.
\end{align}
This leads to the requirement
\begin{align}
T>T_\gc,\quad T_\gc\equiv\frac{\MPl^4}{C^2 m_\bh^2 m}.
\label{eqn:Tgc}
\end{align}
Another requirement is, of course, the gravitational capture of monopoles by PBHs
can be effective only when the PBHs have not evaporated yet. Consider a PBH with mass
$m_\bh$, its ``lifetime'' $\tau_\bh$ can be parameterized as
\begin{align}
\tau_\bh=\varepsilon\frac{m_\bh^3}{\MPl^4}.
\label{eqn:vedef}
\end{align}
For a non-rotating PBH with a negligible charge to mass ratio, we have~\cite{Hooper:2020otu,Gehrman:2022imk}
\begin{align}
\varepsilon=\frac{10240\pi}{\mathcal{G}\langle g_{\star,H}\rangle},
\end{align}
with $\mathcal{G}\simeq 3.8$ is the grey body factor, and $\langle g_{\star,H}\rangle\simeq\mN$ depends
on the particle physics model. Assuming radiation domination, the temperature $T_\ev$ of the universe
at PBH evaporation can be estimated from
\begin{align}
\tau_\bh=\frac{1}{2H(T_\ev)},
\end{align}
which leads to
\begin{align}
T_\ev=(2\varepsilon K)^{-1/2}\bigg(\frac{m_\bh}{\MPl}\bigg)^{-3/2}\MPl.
\label{eqn:Tev}
\end{align}
The gravitational capture of monopoles by PBHs can thus be effective only when
\begin{align}
T_s<T<T_t,\quad T_s\equiv\max\{T_\ev,T_\gc\},\quad T_t\equiv\min\{T_c,T_b\},
\label{eqn:TsTt}
\end{align}
where $T_b$ is the temperature of the universe when the PBHs form.

\subsection{Drift modelling with a monochromatic PBH mass function}

In this work we will focus on a monochromatic PBH mass function with a fixed
$m_\bh$. It is then possible express $n_\bh$ in terms of the energy density fraction
of PBHs $\beta$ at the time of PBH formation as
\begin{align}
(n_\bh m_\bh)|_{\text{formation}}=\beta K_1 T_b^4,
\label{eqn:betadef}
\end{align}
with $T_b$ being the temperature of the plasma at PBH formation. Thus at any temperature $T$ we have
\begin{align}
n_\bh m_\bh =\beta K_1 T_b T^3.
\end{align}

$T_b$ can be related to $m_\bh$ in the following manner. The PBH mass can be related to the Hubble
parameter at formation~\cite{Sasaki:2018dmp}
\begin{align}
m_\bh=\frac{\gamma}{2G}H_\text{form}^{-1},\quad H_\text{form}^{-1}=K\frac{T_b^2}{\MPl},
\label{eqn:gammadef}
\end{align}
Typically $\gamma\simeq 0.2$ \cite{Carr:1975qj} which is the ratio between the PBH mass and the horizon mass at formation.
$T_b$ can then be expressed as
\begin{align}
T_b=\bigg(\frac{\gamma}{2K}\bigg)^{1/2}\bigg(\frac{m_\bh}{\MPl}\bigg)^{-1/2}\MPl.
\label{eqn:Tb}
\end{align}
We then obtain using Eq.~\eqref{eqn:fe1}
\begin{align}
F=K_1 \bigg(\frac{\gamma}{2K}\bigg)^{1/2}C^{-1}\delta\times \beta\bigg(\frac{m_\bh}{\MPl}\bigg)^{-1/2}
\bigg(\frac{T_c}{\MPl}\bigg) T.
\end{align}

Now we return to the evolution equation \eqref{eqn:nmdrift}. It is straightforward to transform
it into a differential equation for $r(T)$ as follows
\begin{align}
\frac{dr}{dT}=\frac{\Delta}{T^2}r^2+\frac{\Phi}{T^2}r,
\label{eqn:rtevolve}
\end{align}
where $\Delta$ is still given by Eq.~\eqref{eqn:delta}, while $\Phi$ is defined as
\begin{align}
\Phi\equiv\frac{\MPl F}{KT}.
\end{align}
It is interesting to note that for a monochromatic PBH mass function, $\Phi$ is also a temperature-independent
quantity of mass dimension $1$. Eq.~\eqref{eqn:rtevolve} can then be solved analytically. For $j=1,2$, suppose at temperature $T_j$, the yield $r$ is $r_j$. Then if $\Phi=0,\Delta>0$ (i.e. only monopole annihilation is effective),
the solution to Eq.~\eqref{eqn:rtevolve} is still given by Eq.~\eqref{eqn:r1r2ann}. When $\Phi>0,\Delta>0$, we introduce
\begin{align}
r_\text{cr}\equiv\frac{\Phi}{\Delta},
\end{align}
and
\begin{align}
\bar{\Phi} \equiv\frac{\Phi}{T_c}.
\end{align}
The solution is then ($z_j=T_j/T_c$, $j=1,2$)
\begin{align}
r_2=\Bigg\{\bigg(\frac{1}{r_\text{cr}}+\frac{1}{r_1}\bigg)\exp\bigg[\bar{\Phi}\bigg(
\frac{1}{z_2}-\frac{1}{z_1}\bigg)\bigg]-\frac{1}{r_\text{cr}} \Bigg\}^{-1}.
\label{eqn:r1r2df}
\end{align}
A special case is when $\Delta=0,\Phi>0$. The solution in this case can be obtained by
simply letting $r_\text{cr}\rightarrow\infty$ in Eq.~\eqref{eqn:r1r2df}. The resulting expression
is simple:
\begin{align}
r_2=r_1\exp\bigg[-\bar{\Phi}\bigg(\frac{1}{z_2}-\frac{1}{z_1}\bigg)\bigg].
\label{eqn:r1r2Phi}
\end{align}

\subsection{Flux description of the drift modelling}
\label{subsec:flux}

In the previous subsections two ways of modelling the monopole capture by PBHs are presented,
however some steps involved may look ad hoc, or lead to questions that remain to be explained. For example:
\begin{enumerate}
\item Eq.~\eqref{eqn:sgvm} is derived in a non-diffusive context. Does it apply here and why should we
use a random walk velocity as discussed previously for $v_M$?
\item In the drift modelling, the drift velocity is a function of the monopole-PBH distance. Why is it appropriate
to use $n_\bh^{-1/3}$ as the typical separation and determine the typical capture time $\tau_\gc$? Does this
treatment still apply if $\tau_\gc>H^{-1}$?
\item Which way of modelling is appropriate?
\end{enumerate}
Motivated by these questions, in this section we present a more intuitive account of the SF modelling and the drift modelling,
justifying our preference for drift modelling in this work.

In the current discussion a monopole in the primordial plasma is subject to three types of forces: gravitational attraction
from nearby PBHs, magnetic attraction/repulsion from nearby monopoles/antimonopoles, and frequent collisions by surrounding
electrically charged particles. To simplify discussion we neglect the mutual influence between monopole annihilation and
capture by PBHs, that is, either process can be studied independently. Here we focus on monopole capture by PBHs, so we drop
the magnetic attraction/repulsion from nearby monopoles/antimonopoles. In the remaining two types of forces, gravitational attraction from nearby PBHs is given by the usual inverse-square law, which for simplicity we consider only the effect of the nearest PBH from the monopole under consideration. The effect of the frequent collisions by surrounding electrically charged
particles is completely analogous to the force experienced by a particle that undergoes Brownian motion, which can be modelled
as the sum of a drag force $\textbf{F}_\text{drag}$ that is approximately proportional to $-\textbf{v}$ ($\textbf{v}$ is the monopole velocity) and a
rapidly fluctuating force $\textbf{F}_\text{fl}(t)$~\cite{Pathria:2022hda}. $\textbf{F}_\text{fl}(t)$ averages to zero over long intervals of time.\footnote{By ``long intervals of time'' we mean when compared to $\tau^\star$ which is the characteristic time between two collisions with electrically charged particles. Note $\tau_M$ introduced in Eq.~\eqref{eqn:taum} which we call the
mean free time is conceptually different from and numerically much larger than $\tau^\star$.} For non-relativistic monopoles,
we may use Newtonian mechanics to describe its motion:
\begin{align}
m\frac{d\textbf{v}}{dt}=-\frac{m_\bh m}{\MPl^2}\frac{1}{\bar{R}^2}\bar{\textbf{R}}+\textbf{F}_\text{drag}+\textbf{F}_\text{fl}(t),
\quad \overline{\textbf{F}_\text{fl}(t)}=0,
\label{eqn:newtonmotion}
\end{align}
Here $\bar{\textbf{R}}$ is the distance vector that points from the PBH to the monopole, and $\overline{\textbf{F}_\text{fl}(t)}$
denotes the average of $\textbf{F}_\text{fl}(t)$ over long intervals of time. Then if we take the ensemble average
of Eq.~\eqref{eqn:newtonmotion}, we obtain ($\langle\textbf{v}\rangle$ denotes the ensemble average of $\textbf{v}$)\footnote{By ``ensemble average'' we are implicitly consider a large number of systems similar to the one originally under consideration
and studying their motion in a \emph{statistical} sense. That is, the initial positions and velocities of the monopoles and PBHs and the macroscopic proerties of the primordial plasma are the same among ensemble members, but the microscopic properties of the primordial plasma differ from member of member.}
\begin{align}
m\frac{d\langle\textbf{v}\rangle}{dt}=-\frac{m_\bh m}{\MPl^2}\frac{1}{\bar{R}^2}\bar{\textbf{R}}-CT^2\langle\textbf{v}\rangle,
\label{eqn:eamotion}
\end{align}
For definiteness we have plugged in Eq.~\eqref{eqn:dragforce} for the drag force (setting $f=1$), and used the fact that
$\langle\textbf{F}_\text{fl}(t)\rangle=0$ by the very nature of $\textbf{F}_\text{fl}(t)$. Regardless of the initial velocity, eventually $\langle\textbf{v}\rangle$ will tend to its equilibrium value, determined from
the balancing between the gravitational attraction and the drag force. Nevertheless, this does not mean the monopole velocity
of all ensemble members will eventually be the same. This is because we are considering the monopole motion in a statistical sense, while the ensemble-averaged equation Eq.~\eqref{eqn:eamotion} only constrains the first moment of the random variable $\textbf{v}$.
For example, a simpler case would be if there is no PBH, so the gravitational attraction does not exist. Then the equilibrium
value of $\langle\textbf{v}\rangle$ will be zero. Of course this does not mean monopoles in all ensemble members will sit still.
In any case there is a non-zero temperature $T$, thus in thermal equilibrium monopoles must exhibit a root-mean-square velocity
that satisfy $\langle\textbf{v}^2\rangle=3T/m$, which is related to the second moment of $\textbf{v}$.

Therefore, while the evolution of the first moment of $\textbf{v}$ is determined by the competition between the gravitational attraction and the drag force,the evolution of the second moment of $\textbf{v}$ is further controlled by $\textbf{F}_\text{fl}(t)$. If the gravitational attraction is negligible, which is the case for very large monopole-PBH distance
$\bar{R}$, then the equilibrium value of $\langle\textbf{v}\rangle$ is zero. For such a monopole with an initial thermal velocity
of $v_T=(3T/m)^{1/2}$, the drag force will make it forget this initial velocity in a characteristic time $\tau_M$ (see Eq.~\eqref{eqn:taum}), however $\textbf{F}_\text{fl}(t)$ will boost the monopole velocity to $\sim(3T/m)^{1/2}$ during a time
interval of order $\sim\tau_M$ to maintain $\langle\textbf{v}^2\rangle=3T/m$ (though the direction of the velocity may change randomly). It is in this sense that $\tau_M$ can be viewed as the characteristic time for the monopole to change its velocity
significantly, and $\ell=v_T\tau_M$ can be viewed as the monopole mean free path.

For small gravitational attraction, the equilibrium value of $\langle\textbf{v}\rangle$ (the drift velocity) is small but nonzero. In such a case, if we trace the motion of the monopole, then in short periods of time (but still larger than $\tau_M$) it would be
very difficult to tell the influence of the drift velocity. That is, the motion of the monopole is thermally dominated (i.e.
dominated by $\textbf{F}_\text{fl}(t)$). For large gravitational attraction, however, the monopole drift velocity is large,
and the motion of the monopole is gravitationally dominated, while the thermal effect only leads to relatively small corrections.
The boundary value of $\bar{R}$ between small and large gravitational attraction regimes can be estimated by requiring
\begin{align}
\frac{m_\bh m}{\MPl^2}\frac{1}{\bar{R}}=T,
\label{eqn:bdrbar}
\end{align}
That is, the amount of gravitational potential energy is equal to the thermal energy of the monopole. The solution of
Eq.~\eqref{eqn:bdrbar} is just the gravitational capture radius $r_c^\gc$ introduced in Eq.~\eqref{eqn:rcgc}.

The gravitational capture of monopoles by PBHs can be divided into two regimes, according to whether
$r_c^\gc>\ell$ or $r_c^\gc<\ell$. If $r_c^\gc>\ell$, then for a monopole satisfying $\bar{R}<r_c^\gc$, its motion
is dominated by the drifting towards the PBH, and the probability for its thermal motion (characterized by $\ell$)
to successfully counteract the drifting and kick it away from the PBH is small (since $\ell<r_c^\gc$). Thus, once a monopole
is inside the region $\bar{R}<r_c^\gc$, its fate is doomed. This $r_c^\gc>\ell$ regime is called diffusive capture.
On the other hand, if $r_c^\gc<\ell$, then even if a monopole satisfies $\bar{R}<r_c^\gc$, there still can be
a quite large probability that its thermal motion may successfully counteract the drifting and kick the monopole away
from the PBH since $\ell>r_c^\gc$. This $r_c^\gc<\ell$ regime is called non-diffusive capture. For a monochromatic
PBH mass function, which is the case studied in this work, $r_c^\gc\propto T^{-1}$ while $\ell\propto T^{-3/2}$,
diffusive capture is effective for $T>T_\gc$, with $T_\gc$ given in Eq.~\eqref{eqn:Tgc}. The SF paper ~\cite{Stojkovic:2004hz}
however considers an extended PBH mass function such that the ratio between the PBH mass and the horizon mass at any given
time is effectively a constant, leading to a situation in which the diffusive capture starts at some high temperature
and practically never ends.

Based on the above physical picture, we now demonstrate that the drift modelling introduced in Sec.~\ref{sec:gc} can also be
reformulated in a flux language. In the diffusive capture regime, the reasonable candidate for the cross section should
be
\begin{align}
\sigma_{g\text{D}}\equiv\pi (r_c^\gc)^2,
\label{eqn:sigmagD}
\end{align}
since once a monopole is inside the region $\bar{R}<r_c^\gc$, its fate is doomed. The appropriate candidate for the flux velocity
should be the drift velocity at $\bar{R}=r_c^\gc$, which is found to be
\begin{align}
v_{M\text{D}}\equiv u_\text{D}(r_c^\gc)=\frac{m_\bh m}{\MPl^2}\frac{1}{CT^2(r_c^\gc)^2}=\frac{\MPl^2}{Cmm_\bh},
\label{eqn:incident}
\end{align}
Then the capture term coefficient $F$ in Eq.~\eqref{eqn:nmdrift} can be expressed as, in the flux description
\footnote{Here we neglect the potential inhomogeneity of $n_M$ during evolution, as in the parameter region considered
in this work, one may verify that $\frac{r_c^\gc}{n_\bh^{-1/3}}\ll 1$.}
\begin{align}
F=\sigma_{g\text{D}}v_{M\text{D}}n_\bh=\pi \frac{n_\bh m_\bh m}{\MPl^2 CT^2},
\label{eqn:Fpi}
\end{align}
Apart from an $\mathcal{O}(1)$ factor, this expression coincides with Eq.~\eqref{eqn:fe1} found in the drift modelling
derivation. Therefore the drift modelling with the typical drift velocity $u_\text{D}(n_\bh^{-1/3})$ can be justified
by the flux description with a capture cross section $\sigma_{g\text{D}}$ and an incident monopole velocity
$v_{M\text{D}}$.

In the drift modelling the typical capture time $\tau_\gc=F^{-1}$ (see Eq.~\eqref{eqn:captime}) can become
larger than the Hubble time $t=\frac{1}{2H}$, raising concerns about the self-consistency of the treatment.
Nevertheless, this issue is fixed in the flux description. In the flux description, we may define a characteristic
timescale
\begin{align}
\tau_\text{cap}\equiv\frac{r_c^\gc}{u_\text{D}(r_c^\gc)}=\frac{Cm^2m_\bh^2}{\MPl^4 T},
\end{align}
which represents the typical time needed for a monopole to get into the PBH once it is inside the capture radius.
The analysis based on the flux description is valid as long as $\tau_\text{cap}<H^{-1}$, that is
\begin{align}
\tau_\text{cap}H<1,
\end{align}
which reads in terms of reduced variables ($x\equiv\frac{T_c}{\MPl},y\equiv\frac{m_\bh}{\MPl},z\equiv\frac{T}{T_c}$)
\begin{align}
CK\delta^2x^3 y^2 z<1,
\label{eqn:comp1}
\end{align}
On the other hand, the defining condition in the diffusive capture regime $r_c^\gc>\ell$ becomes
\begin{align}
C\delta^{1/2}xyz^{1/2}>1,
\label{eqn:comp2}
\end{align}
The compatibility between Eq.~\eqref{eqn:comp1} and Eq.~\eqref{eqn:comp2} only requires
\begin{align}
K\delta x<C,
\end{align}
which is always satisfied for the value of $x$ under consideration in this work. Thus $\tau_\text{cap}<H^{-1}$ is always
satisfied, which in turn justifies the self-consistency of the drift modelling even if $\tau_\gc>H^{-1}$.

The flux description now allows a more direct comparison between the two ways of modelling of the monopole capture
by PBHs. In fact, apart from an $\mathcal{O}(1)$ factor, the incident velocity of the drift modelling obtained in Eq.~\eqref{eqn:incident} is just incident velocity $v_M$ used in the SF modelling, c.f. Eq.~\eqref{eqn:incidentSF}.
Therefore, when both ways of modelling are framed in a flux description, we see that the same incident velocity is
used. So the difference between them essentially comes from the capture cross section. In the SF modelling, the capture
cross section is taken to be $\sigma_g$ (see Eq.~\eqref{eqn:sgvm} and Eq.~\eqref{eqn:sigmag}), while in the drift modelling,
the effective capture cross section is $\sigma_{g\text{D}}$ (see Eq.~\eqref{eqn:sigmagD}). Their ratio is found to be
\begin{align}
\frac{\sigma_g}{\sigma_{g\text{D}}}=\frac{16}{3}C^2\frac{m_\bh^2 T^2}{\MPl^4}=\frac{16}{3}C^2 x^2 y^2 z^2,
\end{align}
Thus for a given PBH mass, at high temperature we have $\frac{\sigma_g}{\sigma_{g\text{D}}}\gg 1$, while at low
temperature it is possible to have $\frac{\sigma_g}{\sigma_{g\text{D}}}<1$.

From the physical picture of diffusive capture described above, it should be clear that the appropriate capture
cross section to be used is $\sigma_{g\text{D}}$ rather than $\sigma_g$. The region $\bar{R}<r_c^\gc$ (corresponding to $\sigma_{g\text{D}}$) acts as an effective extended event horizon of the PBH for the monopole, since once a monopole gets inside, the probability
for it to escape is practically small. On the other hand, $\sigma_g$ is derived in a non-diffusive setting (see Chapter 7 of
Ref.~\cite{Frolov:2011bhp}), by solving the geodesic trajectory of a test massive particle in a Schwarzschild geometry and
finding the critical impact parameter that leads to gravitational capture. Such a computation does not make sense
if the motion of the test particle is also seriously affected by collisions with other particles along the journey.
In the current monopole capture problem, this means $\sigma_g$ does not make sense if it exceeds $\pi \ell^2$, since $\ell$
roughly defines the maximum distance of the monopole motion that is not seriously affected by collisions with plasma particles.\footnote{There is a similar reasoning used in determining the integration limits in calculating the parameter $C$
that characterizes strength of the drag force; see discussion on page 404 of Ref.~\cite{Vilenkin:2000jqa}.}
However, in the SF modelling $\sigma_g>\pi(r_c^\gc)^2>\pi \ell^2$ in most of the temperature range, and thus the use of
$\sigma_g$ as the effective capture cross section is problematic.

\section{Magnetic charge fluctuation}
\label{sec:mcf}

There are two types of magnetic charge fluctuation regarding magnetically charged PBHs
formed in the early universe. First, even if a PBH is formed in a magnetically neutral
manner, it may capture monopoles and antimonopoles with fluctuating numbers, resulting in
some residual magnetic charge. Second, if a PBH is formed after the symmetry breaking
phase transition, it is likely that at formation a horizon volume contains monopoles
and antimonopoles with fluctuating numbers, resulting in a net magnetic charge carried
by the PBH already at formation. We discuss these two types of magnetic charge fluctuation
in turn.

\subsection{Magnetic charge fluctuation from monopole capture}

We first estimate the total number of monopoles and antimonopoles (i.e. regardless of the sign of
the magnetic charge) captured by each PBH on average. The drift modelling and a monochromatic
PBH mass function is assumed. To simplifying the expressions, we will use the reduced variables
$z\equiv T/T_c$ and
\begin{align}
y\equiv\frac{m_\bh}{\MPl}.
\end{align}
The value of $r$ reduced due to gravitational capture between $T_t$
and $T_s$ is (see Eq.~\eqref{eqn:rtevolve})
\begin{align}
\kappa\equiv\Phi\int_{T_s}^{T_t}r(T)T^{-2}dT,
\end{align}
here $r(T)$ is the solution of Eq.~\eqref{eqn:rtevolve}, keeping in mind the temperature ranges
in which monopole annihilation and/or capture by PBHs are effective. In terms of the
reduced variables $z_s\equiv T_s/T_c,z_t\equiv T_t/T_c,\bar{\Phi}=\Phi/T_c$, $\kappa$ can be expressed as
\begin{align}
\kappa=\bar{\Phi}\int_{z_s}^{z_t} r(z)z^{-2}dz.
\end{align}
The average number of monopoles (including antimonopoles) captured by each PBH is then
\begin{align}
n_2=\kappa\frac{s(T_t)}{n_\bh(T_t)},
\end{align}
which is computed to be
\begin{align}
n_2=\frac{1}{3}\bigg(\frac{\pi\mN}{5}\bigg)^{1/2}C^{-1}\delta y\times\int_{z_s}^{z_t} r(z)z^{-2}dz.
\label{eqn:n2e1}
\end{align}
The residual magnetic charge obtained by the PBH is
\begin{align}
\chi_\gc=\chi\sqrt{n_2}.
\label{eqn:chigc}
\end{align}
Eq.~\eqref{eqn:chigc} is based on the assumption that for each PBH, monopoles and antimonopoles are
captured independently. This assumption breaks down when a PBH already obtains a large residual magnetic
charge of $+\chi_\bh$, since then it is preferable for it to capture a monopole with charge $-\chi$ instead
of $+\chi$. If we require the gravitational force between it and a monopole with charge $+\chi$ be larger than
the corresponding magnetic force, then this sets a bound on $\chi_\bh$~\cite{Ghosh:2020tdu}, which can be expressed
in terms of reduced variables as
\begin{align}
\chi_\bh\lesssim\chi_\bh^{\text{lim}},\quad \chi_\bh^{\text{lim}}\equiv\delta\chi^{-1}g^{-2}xy.
\end{align}

\subsection{Magnetic charge fluctuation at PBH formation}

If PBHs were formed after the production of magnetic monopoles, magnetic charge fluctuation
in a horizon volume at formation may already lead to the formation of PBH with
an initial magnetic charge.\footnote{For reviews and discussions on PBHs and their production mechanism, see e.g. Refs.~\cite{Khlopov:2008qy,Calmet:2015fua,Sasaki:2018dmp,Carr:2020gox,Carr:2020xqk,Villanueva-Domingo:2021spv,Carr:2021bzv,
Escriva:2022duf,Liu:2021svg,Choudhury:2013woa,Choudhury:2023vuj,Choudhury:2023jlt}.} This mechanism is similar to the formation mechanism
of primordial dark extremal black holes as discussed in Ref.~\cite{Bai:2019zcd}. The only
difference is that the number density of magnetic monopoles must be obtained as a solution
of the evolution equation \eqref{eqn:rtevolve}.

At $T=T_b<T_c$ the expected number of monopoles (or antimonopoles) per horizon volume is
\begin{align}
\langle N_\col\rangle\simeq\frac{4\pi}{3}n_M(T_b)H_\text{form}^{-3}.
\end{align}
Using
\begin{align}
n_M(T_b)=r(T_b)s(T_b),\quad H_\text{form}=K\frac{T_b^2}{\MPl},
\end{align}
$\langle N_\col\rangle$ can be expressed as (in terms of reduced variables)
\begin{align}
\langle N_\col\rangle\simeq\frac{4\pi}{3}r(z_b)K_2 K^{-3/2}\bigg(\frac{\gamma}{2}\bigg)^{-3/2}y^{3/2}.
\label{eqn:Ncol1}
\end{align}
For a monochromatic PBH mass function, between $T_c$ and $T_b$ only monopole annihilation is effective.
$r(z_b)$ is given by
\begin{align}
r(z_b)=\bigg[\frac{1}{r_i}+\frac{\bar{\Phi}}{r_\text{cr}}\bigg(\frac{1}{z_b}-1\bigg)\bigg]^{-1}.
\label{eqn:rzb1}
\end{align}
If the numbers of monopoles and antimonopoles follow independent Poisson distributions,
then the resulting magnetic charge distribution has a mean value zero and a standard deviation of~\cite{Bai:2019zcd}
\begin{align}
\chi_\col=\chi\sqrt{(2\langle N_\col\rangle)},
\end{align}
thus it is typical to have an initial magnetic charge of $\chi_\col$ when the PBHs form.

We emphasize here that the above treatment of magnetic charge fluctuation
at PBH formation is oversimplified, as it neglects the correlation between monopoles and antimonopoles, and
also potential correlation between energy density perturbation and charge asymmetry fluctuation~\cite{Bai:2019zcd}.

We note that magnetic charge fluctuation at PBH formation is also considered in Ref.~\cite{Araya:2020tds} in which
magnetic PBHs are proposed to be responsible for the generation of cosmic magnetic fields. The difference from
our scenario is that in Ref.~\cite{Araya:2020tds} the magnetic PBHs do not evaporate significantly prior to BBN,
and thus they will neither evolve to near-extremal magnetic black holes, nor evaporate completely. Instead, they
are born as magnetic PBHs with a tiny magnetic charge-to-mass ratio.

\subsection{The issue of cosmological stability}

The residual magnetic charge obtained by a PBH, either from its formation or from monopole capture,
may have a large impact on its fate. If a PBH carries some magnetic charge, it is unstable against
emitting magnetic monopoles via pair creation or breaking into smaller magnetic black holes.\footnote{This is related to the weak gravity conjecture; see e.g. Refs.~\cite{Arkani-Hamed:2006emk,Cheung:2014vva,Heidenreich:2019zkl,Aharony:2021mpc,Antipin:2021rsh}.} For a near-extremal magnetic black hole with
magnetic charge $\chi_\bh$, its lifetime can be roughly estimated as~\cite{Khriplovich:2002qn,Khriplovich:1999gm}
\begin{align}
\tau_{\text{mbh}}\sim \MPl^{-1}\exp\bigg(\frac{m^2}{\MPl^2}\pi g\chi_\bh\bigg),
\end{align}
so the requirement of cosmological stability ($\tau_\bh\gtrsim 10^{18}\,\text{s}$) translates into
\begin{align}
\chi_\bh\gtrsim10^{-2}x^{-2}.
\end{align}
For example, if $x=10^{-4}$, this requires $\chi_\bh\gtrsim 10^6$~\cite{Maldacena:2020skw}.

\section{Preparation for the analysis}
\label{sec:prep}

In the previous sections we described the modelling of the key physical processes in this study, namely monopole annihilation and gravitational capture of monopoles by PBHs. We also outline the computation of two types of
magnetic charge fluctuations. In this section we set the stage for concrete analyses to be performed in Sec.~\ref{sec:abp}.

\subsection{Particle physics scenarios}

Magnetic monopoles are generic predictions of grand unified or partially unified gauge theories.
For definiteness, in the analysis we consider monopoles coming from a Pati-Salam gauge theory
with Pati-Salam breaking scale between $10^{10}\GeV$ and $10^{15}\GeV$\footnote{For early study of magnetic monopoles related to Pati-Salam models, see Refs.~\cite{Lazarides:1980va,Lazarides:1980cc}.}. The main reasons are:
\begin{enumerate}
\item In contrast to grand unified theories (GUT), partially unified theories such as the Pati-Salam
extensions of the SM allows for a more flexible symmetry breaking scale, ranging from just below
the Planck scale, to as low as $\mathcal{O}(10\TeV)$ depending on the field content~\cite{Hartmann:2014fya,DiLuzio:2020xgc,Dolan:2020doe,Cacciapaglia:2019dsq,Cacciapaglia:2020jvj}.
\item The nonobservation of tensor perturbations in the CMB requires the reheating temperature
$T_\text{RH}$ to satisfy $T_\text{RH}\lesssim 10^{16}\GeV$, thus GUT monopoles (associated with unification scales
higher than $T_\text{RH}$ are likely to have already been diluted sufficiently by inflation~\cite{Lozanov:2019jxc}. This is not necessarily the case for Pati-Salam monopoles.
\item As was mentioned in Sec.~\ref{sec:Intro}, we consider a symmetry breaking scale of $\gtrsim 10^{10}\GeV$ in order
to avoid complication from reproducing monopoles from cooling of hot spots created during PBH evaporation~\cite{Das:2021wei,He:2022wwy}.
\end{enumerate}

In Pati-Salam models, gauge symmetry breaking steps can be written as
\begin{align}
&\quad SU(4)_\text{PS}\times SU(2)_L\times SU(2)_R \nonumber, \\
&\rightarrow SU(3)_c\times SU(2)_L\times SU(2)_R\times U(1)_{B-L} \nonumber, \\
&\rightarrow SU(3)_c\times SU(2)_L\times U(1)_Y \nonumber, \\
&\rightarrow SU(3)_c\times U(1)_{EM}.
\end{align}
Monopoles already arise in the first step of breaking as $U(1)_{B-L}$ monopoles that
also carry non-Abelian magnetic charges. The $U(1)_{B-L}$ monopoles survive
the next several stages of symmetry breaking and evolve into
electromagnetic magnetic monoples with also color magnetic charge.
In Pati-Salam models, the minimal magnetic monopoles carry two units of magnetic
charge, corresponding to $\chi=2$ in our notation. Their mass is
\begin{align}
m=\frac{4\pi M_\text{PS}}{g_\text{PS}},
\end{align}
with $M_\text{PS}$ being the Pati-Salam symmetry breaking scale (which will be identified as
$T_c$), and $g_\text{PS}$ being the Pati-Salam gauge coupling.

Thus in the early universe the Pati-Salam monopoles may correspond to $U(1)$ generators
different from $U(1)_{EM}$, and also carry non-Abelian magnetic charges that could lead to
long-range forces. The monopole's non-Abelian magnetic charge is expected to be affected
frequently by surrounding plasma particles and the associated force is expected to be averaged
to zero. This leaves the long-range force associated with $U(1)$'s uncancelled. We will neglect
the complicated evolution of the $U(1)$ coupling due to running, matching, and breaking and just
use $g$ defined in Eq.~\eqref{eqn:unitmag} for the unit magnetic charge, keeping in mind that
$\mathcal{O}(1)$ uncertainties in the value of $g$ is expected due to this approximation.

Though the Pati-Salam model is taken as the benchmark scenario in our study, it is straightforward
to extend the analysis to other models as one only needs to change the relevant parameters accordingly.

\subsection{Parameters and variables}

For convenience, parameters that are relevant to the analysis
are summarized in Table~\ref{tab:parameters}. ``Reference point value'' indicates
the values adopted in numerical results, while ``Floating range'' indicates
the range in which the parameters are allowed to vary taking into account of
uncertainties and the need to consider alternative scenarios.\footnote{``Floating range'' is shown for
illustration purposes but are not used in numerical analysis} Parameters $K$, $K_1$, and $K_2$
frequently appear in the equations, but they are just functions of $\mN$, defined
in Eqs.~\eqref{eqn:k} and \eqref{eqn:k1k2}. Several reduced (dimensionless) variables
that appear in the analysis are
\begin{align}
r\equiv\frac{n_M}{s},\quad\quad z\equiv\frac{T}{T_c},
\end{align}
and
\begin{align}
x\equiv\frac{T_c}{\MPl},\quad y\equiv\frac{m_\bh}{\MPl},\quad \beta,
\end{align}
recalling that $\beta$ is defined in Eq.~\eqref{eqn:betadef}. Each benchmark point should be
specified by a set of $x,y,\beta$ and the initial value of $r$, which then determines the solution of $r$ as a function
of $z$ by solving Eq.~\eqref{eqn:rtevolve}. The solution $r(z)$ can be employed to compute
the magnetic charge fluctuation and compare against the Parker's bound which requires the
total final yield $r_\text{fi}$ to satisfy~\cite{Murayama:2009nj}\footnote{There are various versions of the Parker's bound based on different assumptions~\cite{Turner:1982ag,Rephaeli:1982nv,Adams:1993fj,Lewis:1999zm,Kobayashi:2022qpl}, and the bounds in fact depend on the monopole mass. These subtleties currently do not affect the discussion in this work.}
\begin{align}
r_\text{fi}\lesssim 10^{-26}.
\end{align}
Note that $r_\text{fi}$ should include contributions from both the unabsorbed magnetic monopoles,
and the magnetic monopoles that are decay products of the evaporating magnetic PBHs.

\begin{table*}[t!]
\begin{center}
\begin{tabular}{|c|c|c|c|}
\hline
Parameter & Definition & Reference point value & Floating range \\
\hline
$\MPl$ & Eq.~\eqref{eqn:Planckmass} & $1.2\times 10^{19}\GeV=2.2\times 10^{-5}\,\text{g}$ & NA \\
\hline
$\mN$ & Eq.~\eqref{eqn:k1k2} & $100$ & $100\lesssim\mN\lesssim 1000$ \\
\hline
$g$ & Eq.~\eqref{eqn:unitmag} & 5.9 & $1\lesssim g\lesssim 10$ \\
\hline
$\chi$ & Eq.~\eqref{eqn:chidef} & 2 & $\chi=1\,\text{or}\,2$ \\
\hline
$\delta$ & Eq.~\eqref{eqn:deltax} & $50$ & $10\lesssim\delta\lesssim 100 $ \\
\hline
$C$ & Eq.~\eqref{eqn:Cdef} & $200$ & $100\lesssim C\lesssim 1000$ \\
\hline
$\gamma$ & Eq.~\eqref{eqn:gammadef} & $0.2$ & $0.01\lesssim\gamma\lesssim 1$ \\
\hline
$\varepsilon$ & Eq.~\eqref{eqn:vedef} & 100 & $10\lesssim\varepsilon\lesssim 300$ \\
\hline\hline
$\nu$ & Eq.~\eqref{eqn:nudef} & $0.5$ & $0.5\lesssim\nu\lesssim 0.8$ \\
\hline
$p$ & Eq.~\eqref{eqn:defp} & $0.1$ & $0.01\lesssim p\lesssim 0.1$ \\
\hline
\end{tabular}
\caption{\label{tab:parameters} Summary of the parameters that appear in the analysis, with their definitions, reference point values, and floating range. In the last two rows, $\nu$ and $p$ only affect the initial value of $r$.}
\end{center}
\end{table*}

\subsection{Characteristic temperatures}

Apart from the critical temperature of the symmetry breaking phase transition $T_c$, in the analysis
we encounter four other important characteristic temperatures, which are
\begin{enumerate}
\item $T_b$, the temperature of the universe when the PBHs form; see Eq.~\eqref{eqn:Tb}.
\item $T_\ann$, the temperature above of the universe which monopole annihilation remains effective; see Eq.~\eqref{eqn:Tann}.
\item $T_\ev$, the temperature of the universe when the PBHs evaporate completely; see Eq.~\eqref{eqn:Tev}.
\item $T_\gc$, the temperature of the universe when the gravitational capture of monopoles by PBHs remains
effective (assuming PBHs have not evaporated yet); see Eq.~\eqref{eqn:Tgc}.
\end{enumerate}
For convenience we will often use the corresponding reduced temperatures instead. They are defined as
\begin{align}
z_b\equiv\frac{T_b}{T_c},\quad z_\ann\equiv\frac{T_\ann}{T_c},\quad
z_\ev\equiv\frac{T_\ev}{T_c},\quad z_\gc\equiv\frac{T_\gc}{T_c}.
\end{align}
Their analytic expressions in terms of the reduced variables and parameters, and the corresponding
reference point expressions are listed in Table~\ref{tab:ztable}.

In accordance with Eq.~\eqref{eqn:TsTt}, we also introduce the corresponding reduced temperatures
\begin{align}
z_s\equiv\frac{T_s}{T_c}=\max\{z_\ev,z_\gc\},\quad z_t\equiv\frac{T_t}{T_c}=\min\{1,z_b\}.
\end{align}

\begin{table*}[t!]
\begin{center}
\begin{tabular}{|c|c|c|}
\hline
Reduced temperature & Analytic expression & Reference point expression  \\
\hline
$z_b$ & $\big(\frac{\gamma}{2K}\big)^{1/2}x^{-1}y^{-1/2}$ & $0.078x^{-1}y^{-1/2}$ \\
\hline
$z_\ann$ & $\delta C^{-2}\chi^{-4} g^{-4}$ & $6.4\times 10^{-8}$ \\
\hline
$z_\ev$ & $(2\varepsilon K)^{-1/2}x^{-1}y^{-3/2}$ & $0.017x^{-1}y^{-3/2}$ \\
\hline
$z_\gc$ & $C^{-2}\delta^{-1}x^{-2}y^{-2}$ & $5\times 10^{-7}x^{-2}y^{-2}$ \\
\hline
\end{tabular}
\caption{\label{tab:ztable} Summary of the reduced characteristic temperatures.}
\end{center}
\end{table*}

\subsection{Overview of the parameter space}

We now consider the requirements on the parameters $x$, $y$, and $\beta$ before solving for the $r(z)$ function.
We require the Pati-Salam symmetry breaking scale to be below the maximum reheating temperature $T_\text{RH}^{\text{max}}\simeq 10^{16}\GeV$, and not lower than $10^{10}\GeV$ to avoid reproducing monopoles via cooling of hot spots created during
PBH evaporation. Therefore $x$ is required to be in the range $10^{-9}\lesssim x\lesssim 10^{-3}$.

For a fixed value of $x$, the range of $y$ is subject to the following constraints.
\begin{enumerate}
\item We require PBHs to form after inflation, that is $T_b\lesssim T_\text{RH}^{\text{max}}\simeq 10^{16}\GeV$. This can be translated into the constraint
    \begin{align}
    y\gtrsim\frac{\gamma}{2K}\bigg(\frac{\MPl}{T_\text{RH}^\text{max}}\bigg)^2,
    \end{align}
    which at the reference point reads
    \begin{align}
    y\gtrsim 8.7\times 10^3.
    \label{eqn:y87}
    \end{align}
\item The temperature of the universe at PBH evaporation should be higher than the BBN temperature, that is $T_\ev\gtrsim T_\text{BBN}$. This can be translated into the constraint
    \begin{align}
    y\lesssim (2\varepsilon K)^{-1/3}\bigg(\frac{\MPl}{T_\text{BBN}}\bigg)^{2/3},
    \end{align}
    which for a conservative value $T_\text{BBN}=1\MeV$ reads at the reference point
    \begin{align}
    y\lesssim 3.5\times 10^{13}.
    \end{align}
    This corresponds to $m_\bh\lesssim 7.7\times 10^8\,\text{g}$.
    \item There exists a temperature range such that gravitational capture of monopoles by PBHs is effective. That is, $z_s<z_t$. It turns out that for the $x$ values of our interest, taking into account Eq.~\eqref{eqn:y87}, the most stringent bound comes from $z_\gc<1$, leading to
        \begin{align}
        y>C^{-1}\delta^{-1/2}x^{-1},
        \end{align}
        which reads at the reference point
        \begin{align}
        y>7.1\times 10^{-4}x^{-1}.
        \end{align}
\end{enumerate}

We require the energy density of the PBHs be always smaller than that of radiation. This condition only needs to be checked just prior to PBH evaporation, which leads to the following constraint on $\beta$ for a fixed value of $y$
\begin{align}
\beta y\lesssim (\gamma\varepsilon)^{-1/2},
\label{eqn:betay}
\end{align}
which reads at the reference point\footnote{When the temperature drops below $\simeq 10\MeV$, $\mN$ drops to about $\simeq 10$~\cite{Husdal:2016haj}, one should use a more stringent constraint such as $\beta y\lesssim 0.02$ instead. This is the case for $T_\ev\lesssim 10\MeV$.}
\begin{align}
\beta y\lesssim 0.22.
\end{align}

We summarize the constraints on $x,y,\beta$ at the reference point as follows
\begin{align}
\begin{cases}
10^{-9}\lesssim x\lesssim 10^{-3}, \\
\max\big\{8.7\times 10^3,\,7.1\times 10^{-4}x^{-1} \big\}\lesssim y\lesssim 3.5\times 10^{13}, \\
\beta\lesssim 0.22 y^{-1}.
\end{cases}
\label{eqn:sumconstraints}
\end{align}

To ensure the self-consistency of the calculation, we also require that monopoles do not dominate the energy density of the universe at PBH evaporation. If for $z\simeq z_\ev$ we have $r\simeq r_p$, then the requirement implies
\begin{align}
\frac{r_p}{z_\ev}\lesssim \delta^{-1},
\label{eqn:rpzev}
\end{align}
which will be imposed in the analysis, except when we consider magnetic charge fluctuation at PBH formation.

\subsection{Initial values}

We consider two scenarios for the initial value of $r$, denoted $r_i$.
The first scenario (denoted by ``Kibble'') is a strongly first-order phase transition that makes $r_i$ saturate the Kibble estimate Eq.~\eqref{eqn:kibble}. The second scenario (denoted by ``KZ'') is a second-order phase transition so that
$r_i$ is given by the Kibble-Zurek estimate \eqref{eqn:KZ} at the reference point.
The initial values for representative values of $T_c$ are listed in Table~\ref{tab:ri}
for illustration purposes.

\begin{table*}[t!]
\begin{center}
\begin{tabular}{|c|c|c|c|c|c|}
\hline
$T_c/\GeV$ & $x$ & $y$ lower bound & $y$ upper bound & $r_i$ (Kibble) & $r_i$ (KZ)  \\
\hline
$1.2\times 10^{15}$ & $10^{-4}$ & $8.7\times 10^3$ & $3.5\times 10^{13}$ & $1.3\times 10^{-10}$ & $3.4\times 10^{-5}$ \\
\hline
$1.2\times 10^{14}$ & $10^{-5}$ & $8.7\times 10^3$ & $3.5\times 10^{13}$ & $1.3\times 10^{-13}$ & $3.4\times 10^{-6}$ \\
\hline
$1.2\times 10^{13}$ & $10^{-6}$ & $8.7\times 10^3$ & $3.5\times 10^{13}$ & $1.3\times 10^{-16}$ & $3.4\times 10^{-7}$ \\
\hline
$1.2\times 10^{12}$ & $10^{-7}$ & $8.7\times 10^3$ & $3.5\times 10^{13}$ & $1.3\times 10^{-19}$ & $3.4\times 10^{-8}$ \\
\hline
$1.2\times 10^{11}$ & $10^{-8}$ & $7.1\times 10^4$ & $3.5\times 10^{13}$ & $1.3\times 10^{-22}$ & $3.4\times 10^{-9}$ \\
\hline
$1.2\times 10^{10}$ & $10^{-9}$ & $7.1\times 10^5$ & $3.5\times 10^{13}$ & $1.3\times 10^{-25}$ & $3.4\times 10^{-10}$ \\
\hline
\end{tabular}
\caption{\label{tab:ri} Summary of the initial values.}
\end{center}
\end{table*}

It will be clear in Sec.~\ref{sec:abp} that in the scenarios we consider, a significant reduction of
the monopole yield due to capture by PBHs does not occur. This implies that for Eq.~\eqref{eqn:rpzev}, we may approximate
\begin{align}
r_p\simeq r(z_\ann)=\min\{r_i,r_\star\}.
\end{align}
Note that at the reference point
\begin{align}
r_\star\simeq 3.5\times 10^{-8}x,
\end{align}
while the Kibble estimate reads (see Table~\ref{tab:ri})
\begin{align}
r_i^\text{Kibble}\simeq 1.3\times 10^2 x^3.
\end{align}
Thus if $r_\star>r_i$, we may deduce a bound on $x$
\begin{align}
x\lesssim 1.6\times 10^{-5},
\label{eqn:xKibble}
\end{align}
since the Kibble estimate sets a lower bound on $r_i$.

\subsection{Evolution of $r$}

The monopole annihilation is effective for $z\in[z_\ann,1]$, gravitational capture of monopoles
by PBHs is effective for $z\in[z_s,z_t]$. The evolution of $r$ as a function of $z$
can then be obtained from solving Eq.~\eqref{eqn:rtevolve} and switching to the reduced variables.
The solutions can be collected from Eqs.~\eqref{eqn:r1r2ann}, \eqref{eqn:r1r2df}, and \eqref{eqn:r1r2Phi}.
Previously, a few derived parameters are employed to express the solutions:
\begin{align}
\bar{\Phi}\equiv\frac{\Phi}{T_c},\quad r_\text{cr}\equiv\frac{\Phi}{\Delta}.
\end{align}
We will also introduce
\begin{align}
\bar{\Delta}\equiv\frac{\Delta}{T_c}=\frac{\bar{\Phi}}{r_\text{cr}}.
\end{align}
Using reduced variables, they can be cast into
\begin{align}
\bar{\Phi} & =\bigg(\frac{\gamma}{2}\bigg)^{1/2}K_1 K^{-3/2}C^{-1}\delta\times\beta y^{-1/2}, \\
r_\text{cr} & =\bigg(\frac{\gamma}{2}\bigg)^{1/2}K_1 K_2^{-1} K^{-1/2}\chi^{-2} g^{-2}\delta\times\beta xy^{-1/2}, \\
\bar{\Delta} & =K_2 K^{-1}C^{-1}\chi^2 g^2 x^{-1}.
\end{align}
At the reference point, these equations become
\begin{align}
\bar{\Phi}\simeq 0.038\beta y^{-1/2},\quad r_\text{cr}\simeq 0.021\beta xy^{-1/2},
\quad \bar{\Delta}\simeq 1.84x^{-1}.
\end{align}
The solution to Eq.~\eqref{eqn:rtevolve} can be summarized as ($r_1\equiv r(z_1),r_2\equiv r(z_2)$)
\begin{align}
r_2 &=\Bigg\{\bigg(\frac{1}{r_\text{cr}}+\frac{1}{r_1}\bigg)\exp\bigg[\bar{\Phi}\bigg(
\frac{1}{z_2}-\frac{1}{z_1}\bigg)\bigg]-\frac{1}{r_\text{cr}} \Bigg\}^{-1},\quad z_1,z_2\in [z_\ann,1]\bigcap [z_s,z_t], \\
r_2 &=r_1\exp\bigg[-\bar{\Phi}\bigg(\frac{1}{z_2}-\frac{1}{z_1}\bigg)\bigg],\quad z_1,z_2\in [z_s,z_t]\bigcap[0,z_\ann],\\
r_2 &=\bigg[\frac{1}{r_1}+\bar{\Delta}\bigg(\frac{1}{z_2}-\frac{1}{z_1}\bigg) \bigg]^{-1},\quad z_1,z_2\in [z_\ann,1]\bigcap\big([0,z_s]\bigcup[z_t,\infty]\big).
\end{align}
We assume when both $\Phi$ and $\Delta$ terms are turned off, $r$ remains constant.

\section{Analysis: modelling comparison and magnetic charge fluctuation}
\label{sec:abp}

\subsection{Comparison between two ways of modelling}

In order to compare between the SF modelling and the drift modelling of the monopole capture process,
we note that in the SF modelling, the evolution of $r$ obeys
\begin{align}
\frac{dr}{dT}=\frac{\Delta}{T^2}r^2+\Xi r,\quad \Xi\equiv\frac{F_\SF}{HT}.
\label{eqn:rtSF}
\end{align}
Here $F_\SF$ is given by Eq.~\eqref{eqn:FSF1}. In the following we concentrate on the effect of gravitational
capture by PBHs, and thus drop the $\Delta$ term which characterizes monopole annihilation.
By introducing
\begin{align}
w\equiv-\ln z,\quad \bar{\Xi}\equiv \Xi T_c,
\end{align}
Eq.~\eqref{eqn:rtSF} can be cast into
\begin{align}
-\frac{d\ln r}{dw}=\bar{\Xi}e^{-w},\quad\text{SF},
\label{eqn:rwsf}
\end{align}
while the corresponding equation for the drift modelling is
\begin{align}
-\frac{d\ln r}{dw}=\bar{\Phi}e^w,\quad\text{Drift}.
\label{eqn:rwdr}
\end{align}
For a monochromatic PBH mass function, $\bar{\Xi}$ can be expressed via reduced variables as
\begin{align}
\bar{\Xi}=2^{3/2}(\gamma K)^{1/2}C\delta\times\beta x^2 y^{3/2},
\end{align}
which reads at the reference point
\begin{align}
\bar{\Xi}\simeq 5.2\times 10^4\times\beta x^2 y^{3/2}.
\end{align}
The right-hand side of Eqs.~\eqref{eqn:rwsf} and \eqref{eqn:rwdr} may be regarded as the
\emph{fractional efficiency} of the monopole yield reduction due to capture by PBHs
in two ways of modelling, at a given temperature characterized by $w$.
For a given benchmark point (i.e. fixed $x$, $y$, and $\beta$), we see the fractional efficiency
shows different behavior as a function of $w$ in two ways of modelling, assuming a monochromatic
PBH mass function for both. For the SF modelling, the fractional efficiency scales as $z=e^{-w}$,
indicating an exponential suppression at low temperature. Since in the SF modelling, $v_M$ and $\sigma_g$
are temperature-independent, this exponential suppression can simply be traced to the decrease of
$n_\bh$ with respect to $T$; see Eq.~\eqref{eqn:fsfdef}. For the drift modelling, the
fractional efficiency scales as $z^{-1}=e^w$, indicating an exponential enhancement at low temperature.
This can be traced to the fact that at low temperature, the drag force is reduced, allowing for a
larger drift velocity. This effect eventually overrides the decrease of the efficiency due to the decrease
of $n_\bh$. The drift modelling thus exhibits typical behavior of diffusive capture, as in the case of
monopole annihilation.

It is instructive to check the maximum fractional efficiency of the monopole yield reduction that can be
reached in two ways of modelling. Taking into account the above-mentioned scaling behavior, and
the fact that monopole capture by PBHs is effective when $z\in [z_s,z_t]$, the maximum fractional efficiency
that can be reached in the SF modelling is
\begin{align}
\bar{\Xi}z_t=
\begin{cases}
2^{3/2}(\gamma K)^{1/2}C\delta\times\beta x^2 y^{3/2}, & \quad z_b\geq 1, \\
2\gamma C\delta\times\beta xy, & \quad z_b<1,
\end{cases}
\end{align}
which reads at the reference point
\begin{align}
\bar{\Xi}z_t=
\begin{cases}
5.2\times 10^4\beta x^2 y^{3/2}, & \quad z_b\geq 1, \\
4\times 10^3\beta xy, & \quad z_b<1.
\end{cases}
\end{align}
The maximum fractional efficiency in the drift modelling is
\begin{align}
\bar{\Phi}z_s^{-1}=
\begin{cases}
\frac{1}{4}\big(\frac{\pi\mN}{5}\big)^{1/2}(\gamma\varepsilon)^{1/2}C^{-1}\delta\times\beta xy, &\quad z_\ev\geq z_\gc, \\
\frac{1}{4}\big(\frac{9\mN}{20\pi}\big)^{1/4}\big(\frac{\gamma}{2}\big)^{1/2}C\delta^2\times\beta x^2 y^{3/2}, &\quad z_\ev<z_\gc,
\end{cases}
\end{align}
which reads at the reference point
\begin{align}
\bar{\Phi}z_s^{-1}=
\begin{cases}
2.2\beta xy, &\quad z_\ev\geq z_\gc, \\
7.7\times 10^4\beta x^2 y^{3/2}, &\quad z_\ev<z_\gc.
\end{cases}
\end{align}
Using Eq.~\eqref{eqn:betay}, it is possible to show that
\begin{align}
\bar{\Xi}z_t & \lesssim 2\gamma^{1/2}\varepsilon^{-1/2}C\delta x, \\
\bar{\Phi}z_s^{-1} & \lesssim \frac{1}{4}\bigg(\frac{\pi\mN}{5}\bigg)^{1/2}C^{-1}\delta x.
\end{align}
which read at the reference point
\begin{align}
\bar{\Xi}z_t & \lesssim 9\times 10^2 x, \\
\bar{\Phi}z_s^{-1} & \lesssim 0.5 x.
\end{align}
The maximum values are saturated when $\beta y\simeq 0.22$. We see that the maximum fractional efficiency
that can be achieved in the SF modelling is three orders of magnitude larger than that of
the drift modelling. This partly explains why the monopole density is found to be efficiently reduced
in the SF paper~\cite{Stojkovic:2004hz}. However, for the $x$ values of our interest,
the maximum fractional efficiencies in both ways of modelling are not larger than $1$, implying
that a significant reduction of monopole yield cannot be achieved in this setup, in both the SF
modelling and the drift modelling. Thus to further improve the monopole reduction efficiency,
one should also consider using an extended PBH mass function instead of a monochromatic one,
as proposed in the SF paper. The reason is that for a monochromatic PBH mass function, since the universe is required to be radiation-domianted at the PBH evaporation, the PBH mass density constraint is only saturated at the time
of PBH evaporation, while at earlier times the available room for PBH mass density is not fully utilized.
Since PBHs may evaporate, using an appropriately chosen extended PBH mass function one may fully utilize the available fraction
of energy density to make PBHs and capture monopoles.

For illustrative purposes, let us consider the specific choice of parameters:
\begin{align}
x=10^{-6},\quad y=10^9,\quad \beta=2\times 10^{-10},\quad r_i=1.3\times 10^{-16}.
\label{eqn:parameterchoice}
\end{align}
This value of $r_i$ just saturates the Kibble estimate (see Table~\ref{tab:ri}), and the choice of $\beta$
almost saturates Eq.~\eqref{eqn:betay}. The characteristic reduced temperatures
are computed to be
\begin{align}
z_b\simeq 2.45,\quad z_\ann\simeq 6.4\times 10^{-8},\quad z_\ev\simeq 5.5\times 10^{-10},\quad z_\gc\simeq 5\times 10^{-13}.
\end{align}
$z_b>1$ so the PBH forms before symmetry-breaking phase transition, $z_\ev>z_\gc$ so $z_\gc$ is irrelevant to the analysis.

\begin{figure}[t]
\centering
	\includegraphics[width=0.8\linewidth]{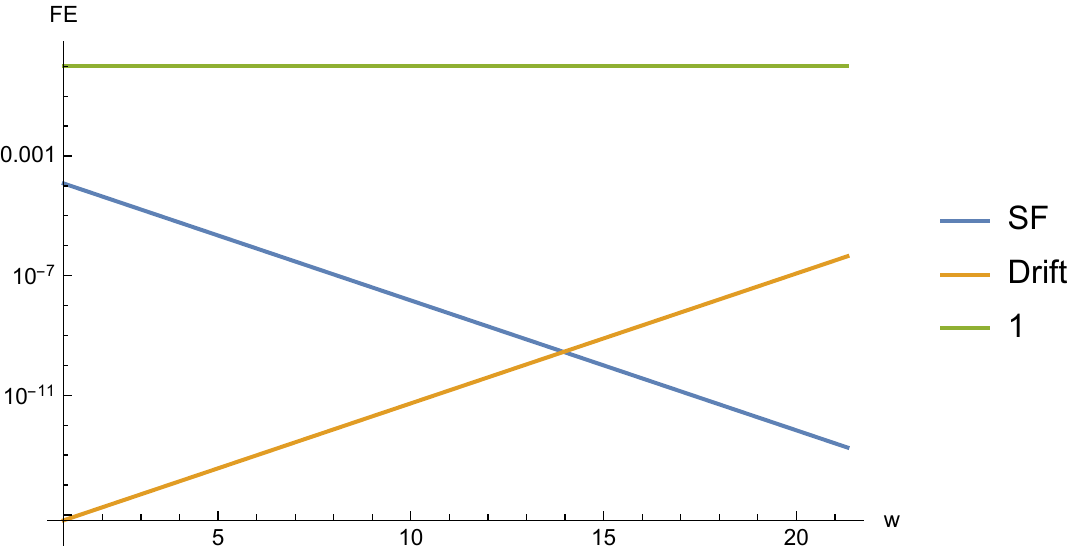}
	\caption{Comparison between the SF modelling and the drift modelling for the fractional efficiency of the monopole yield reduction. Note the vertical axis (the fractional efficiency) is shown on a logarithmic scale.
	}
	\label{fig:BP1A}
\end{figure}

The comparison between the SF modelling and the drift modelling for the fractional efficiency of the monopole yield reduction
is displayed in Fig.~\ref{fig:BP1A} for this set of parameters. The vertical axis indicates the fractional efficiency (FE). The horizontal axis is $w$ and a larger $w$ corresponds to a lower temperature. The display is cut off at the PBH evaporation temperature. It is clear that the drift modelling exhibits a typical behavior of diffusive capture with higher fractional efficiency at the low-temperature end, while the SF modelling exhibits a reversed behavior. With this set of parameters, in both ways of modelling the fractional efficiency turns out to be much smaller than the $-\frac{d\ln r}{dw}=1$ line (the green line in Fig.~\ref{fig:BP1A}). This means in both ways of modelling the monopole capture effect cannot keep up with the cosmic expansion and thus cannot lead to a significant reduction of the monopole yield. Note that this is derived from the assumption of a monochromatic
PBH mass function and radiation domination before PBH evaporation. It is expected that the fractional efficiency can be enhanced once these assumptions are dropped.

For the specific set
of parameters in Eq.~\eqref{eqn:parameterchoice}, the critical value of $z$ that divides the $\frac{\sigma_g}{\sigma_{g\text{D}}}>1$ and $\frac{\sigma_g}{\sigma_{g\text{D}}}<1$ regimes is computed to be
$z\simeq 2.2\times 10^{-6}$. The corresponding value of $w$ is $w=-\ln z\simeq 13$, which just corresponds to
the intersection between the SF line and the Drift line in Fig.~\ref{fig:BP1A}.\footnote{In Fig.~\ref{fig:BP1A} the SF line
and the Drift line intersects at $w\simeq 14$ instead of $w\simeq 13$. This slight discrepancy can be traced to
the factor of $\pi$ in Eq.~\eqref{eqn:Fpi}.} From the above analysis it is clear that the overall higher capture
efficiency achieved in the SF modelling (for a monochromatic PBH mass function) can be traced to the fact that
for most of the temperature range where the diffusive capture is effective, $\frac{\sigma_g}{\sigma_{g\text{D}}}\gg 1$.
Though at low temperature $\frac{\sigma_g}{\sigma_{g\text{D}}}<1$ is achieved, this trend is cut off at $T_\gc$ where
the maximum FE in the drift modelling is not as large as that of the SF modelling.

We also note that with this set of parameters
\begin{align}
r_\star\simeq 3.5\times 10^{-14}>r_i.
\end{align}
Therefore monopole annihilation cannot reduce the monopole yield below $r_i$, either. Since $r_i$ exceeds the Parker's bound by many orders of magnitude, without other mechanisms the monopole problem would remain unresolved. Nevertheless, the analysis demonstrates the differences between two ways of modelling, from which we expect that if the same extended PBH mass function
is used, in the drift modelling it would be harder to solve the monopole problem using PBHs compared to the SF work, and thus
to what extent the monopole problem can be eliminated via capture by PBH should be reexamined.

\subsection{Analysis of magnetic charge fluctuations}

\subsubsection{Monopole capture}

We now consider magnetic charge fluctuation from monopole capture. Approximation to the integral $\int_{z_s}^{z_t} r(z)z^{-2}dz$
that appears in Eq.~\eqref{eqn:n2e1} can be made by noting that the main contribution to the integral
comes from the region where $z$ is close to $z_s$. In that region one may approximate $r(z)\simeq r_p$ as constant, then
\begin{align}
\int_{z_s}^{z_t} r(z)z^{-2}dz\simeq r_p z_s^{-1}\lesssim r_p z_\ev^{-1}\lesssim\delta^{-1},
\end{align}
where we have used Eq.~\eqref{eqn:rpzev} that comes from requiring the monopole energy density be smaller than that of
radiation at PBH evaporation. Then we arrive at an inequality for $n_2$
\begin{align}
n_2\lesssim \frac{1}{3}\sqrt{\frac{\pi\mN}{5}}C^{-1}y,
\end{align}
and the corresponding inequality for $\chi_\gc$ at the reference point
\begin{align}
\chi_\gc\lesssim 0.23 y^{1/2}.
\end{align}
Since $y\lesssim 3.5\times 10^{13}$ (PBHs evaporate before BBN), we find an upper bound on $\chi_\gc$
\begin{align}
\chi_\gc\lesssim 1.4\times 10^6.
\label{eqn:chigc1}
\end{align}
At first sight this suggests the possibility for cosmologically long-lived magnetic black holes
if $x\gtrsim 10^{-4}$. However a closer examination suggests that this is not possible. Basically we need to consider two
cases:
\begin{enumerate}
\item If $r_i<r_\star$, then $r_p=r_i$ with
\begin{align}
n_2\simeq \frac{1}{3}\sqrt{\frac{\pi\mN}{5}}C^{-1}\delta y\times r_i z_\ev^{-1}.
\end{align}
By adjusting the parameters, it is possible to saturate $r_i z_\ev^{-1}\simeq\delta^{-1}$ and thus
achieve $\chi_\gc\simeq 10^6$. However, from Eq.~\eqref{eqn:xKibble} we have $x\lesssim 1.6\times 10^{-5}$,
indicating that $\chi_\gc\simeq 10^6$ is not large enough for the magnetic black hole to be
cosmologically stable.
\item If $r_i>r_\star$, the problem is that the constraint $r_p z_\ev^{-1}\lesssim\delta^{-1}$ becomes
\begin{align}
x^2 y^{3/2}\lesssim 10^4,
\end{align}
at the reference point, which for $x=10^{-4}$ requires $y\lesssim 10^8$, which is far from
the value needed to saturate Eq.~\eqref{eqn:chigc1}. Decreasing $x$ further can only worsen the result.
\end{enumerate}
Thus we conclude that it is not possible to obtain a large enough magnetic charge to make light PBHs
cosmologically stable from monopole capture (at least for parameter choices not very far from our reference points). On one hand, this is related to the drift modelling in which the fractional efficiency is significantly reduced compared to the SF modelling, especially at high temperature. On the other hand, one of the main constraints is due to the requirement that the universe be radiation-dominated up to PBH evaporation.

\subsubsection{Magnetic charge from PBH formation}

To analyze the magnetic charge fluctuation at PBH formation, we consider Eqs.~\eqref{eqn:Ncol1} and \eqref{eqn:rzb1},
and approximate $r(z_b)$ as
\begin{align}
r(z_b)\simeq\min\{r_i,\bar{\Delta}^{-1}z_b\}.
\end{align}
Therefore we may consider two cases, that are $r_i<\bar{\Delta}^{-1}z_b$ and $r_i>\bar{\Delta}^{-1}z_b$, separately.
In both cases we impose the requirement $\chi_\text{col}\gtrsim 10^{-2}x^{-2}$ at the reference point, and also
$y\lesssim 3.5\times 10^{13}$ and $x^2 y\gtrsim 6\times 10^{-3}$ which comes from $z_b<1$. It turns out that in both cases
these considerations lead to
\begin{align}
x\gtrsim 1.8\times 10^{-5},
\end{align}
and it is possible to have a large $\chi_\text{col}$ to obtain a cosmologically stable extremal magnetic black hole.
For example, a benchmark choice can be
\begin{align}
x=10^{-4},\quad y=3.5\times 10^{13},\quad \beta=5.7\times 10^{-16},\quad r_i=6.7\times 10^{-9},
\end{align}
and it turns out
\begin{align}
\chi_\text{col}\simeq 2.2\times 10^7,
\end{align}
which is sufficiently large for $x=10^{-4}$. Since here we are interested in the magnetic charge obtained at
PBH formation, we do not require the monopole energy density to be constrained at the time of PBH evaporation in
this analysis.

\section{Discussion and conclusions}
\label{sec:dnc}

\subsection{Discussion: Non-diffusive monopole capture?}
\label{subsec:anatomy}

In the case of monopole annihilation, it is possible to have non-diffusive capture when $\ell>r_c^{\text{ann}}$,
though its contribution to the reduction of monopole abundance is found to be much smaller than that of diffusive capture.
When it comes to monopole capture by PBHs, is it possible to have non-diffusive capture as well and is its contribution
negligible? The initial expectation is that
once the temperature drops below $T_\gc$, then $r_c^\gc<\ell$, then the effective capture cross section in the non-diffusive
regime should be given by $\sigma_{g\text{ND}}\equiv\min\{\sigma_g,\pi\ell^2\}$. However, there are other constraints
that have to be taken into account. First, we should require $z_\ev<z_\gc$, that is, the PBH evaporation should occur
after the monopole capture enters the non-diffusive regime. Otherwise the PBH evaporation occurs first and the discussion of
non-diffusive capture makes no sense. Using expressions in Table~\ref{tab:ztable} we deduce that $z_\ev<z_\gc$ is equivalent
to
\begin{align}
(2\varepsilon K)^{-1/2}C^2\delta xy^{1/2}<1,
\end{align}
which at the reference point reads
\begin{align}
3.5\times 10^4 xy^{1/2}<1,
\label{eqn:35xy}
\end{align}
On the other hand, the monopole is an extended object with the characteristic length scale of the classical field
configuration being $r_\text{cl}\sim(\alpha m)^{-1}$. If this length scale is larger than the Schwarzschild radius $R_\bh$ of the PBH, then the monopole cannot be viewed as a point particle to be captured. Requiring $R_\bh>r_\text{cl}$
leads to
\begin{align}
\frac{2\alpha m_\bh m}{\MPl^2}>1,
\end{align}
which is approximately equivalent to
\begin{align}
xy>1,
\label{eqn:xy1}
\end{align}
Now let us try to use Eq.~\eqref{eqn:xy1} in Eq.~\eqref{eqn:35xy}. The left-hand side of Eq.~\eqref{eqn:35xy} can be written as
\begin{align}
3.5\times 10^4 xy^{1/2}=3.5\times 10^4 x^{1/2}\times(xy)^{1/2}>3.5\times 10^4 x^{1/2},
\label{eqn:35xy1}
\end{align}
where in the second step we used Eq.~\eqref{eqn:xy1}. However, in this work we restrict ourselves to $x\gtrsim 10^{-9}$
(see Eq.~\eqref{eqn:sumconstraints}), thus Eq.~\eqref{eqn:35xy1} leads to
\begin{align}
3.5\times 10^4 xy^{1/2}>3.5\times 10^4 x^{1/2}>1,
\end{align}
which contradicts the constraint Eq.~\eqref{eqn:35xy}. Therefore, within the parameter range discussed in this work,
to avoid the complication due to the extended field profile of the monopole, we do not consider the non-diffusive capture regime.

Let us comment that besides the size of the classical field profile $r_\text{cl}$, there are two other length scales one
might want to compare with $R_\bh$. One is the Compton wavelength of the monopole, which is $m^{-1}$. However,
$m^{-1}>R_\bh$ does not imply the monopole cannot be captured by PBH. A massless photon by definition has infinite Compton
wavelength but can nevertheless be captured by a black hole. The proper wavelength to be used should be the de Broglie wavelength
for a massive particle (and the optical wavelength for a massless photon)~\cite{Hawking:1971ei}. If we use the thermal
de Broglie wavelength $\lambda_\text{TdB}\sim(mT)^{-1/2}$ for the monopole, gravitational capture by PBH requires
$R_\bh>\lambda_\text{TdB}$, which at the reference point reads $xyz^{1/2}\gtrsim 0.1$, which can be easily violated by
our choice of parameters, such as the specific choice in Eq.~\eqref{eqn:parameterchoice}. Nevertheless, there is an additional twist. As the monopole gets closer to the PBH, its drift velocity increases dramatically and the corresponding de Broglie wavelength becomes much smaller than $\lambda_\text{TdB}$, which can be confirmed by computing $u_\text{D}(\lambda_\text{TdB})$.
It turns out that no matter $u_\text{D}(\lambda_\text{TdB})$ is relativistic or not, the corresponding de Broglie wavelenth
of the monopole at $\bar{R}=\lambda_\text{TdB}$ is smaller than $R_\bh$ and we thus expect the monopole can be captured like
a point particle by the PBH.

\subsection{Conclusions}
\label{subsec:conclusions}

Both magnetic monopoles and black holes are objects with fascinating theoretical properties, and
it is interesting to ask whether they met before in cosmological history, leading to mechanisms
that solve the monopole problem and produce magnetic black holes. In this regard we revisited the
black holes solution to the monopole problem proposed by Stojkovic and Freese~\cite{Stojkovic:2004hz}.
We propose to model monopole capture by PBHs in the same manner as the modelling of monopole annihilation,
which exhibits a typical behavior of diffusive capture. Our drift modelling is compared to the SF one in the case
of a monochromatic PBH mass function, and we find a monopole capture efficiency significantly less
than that of SF modelling. We present an intuitive comparison between the two ways of modelling based on
the flux description and justify our preference for the drift modelling with an appropriate choice of
capture cross section in the diffusive regime. Our result calls for a reanalysis of this black hole solution to the monopole
problem by using an appropriately extended PBH mass function with the drift modelling, which we left for future study.

We have also investigated two types of magnetic charge fluctuation: from PBH formation, or from monopole capture.
We find that if the magnetic charge is acquired at PBH formation, it is possible to make it sufficiently large
such that the resulting extremal magnetic black hole is cosmologically stable. However, if the magnetic charge is acquired
from monopole capture alone, it is not possible to have a sufficiently large residual magnetic charge to
make a cosmologically stable extremal magnetic black hole, due to the assumption that the universe is radiation-dominated
before PBH evaporation.

The analysis done in this work can be extended or refined in a number of directions. Besides using an extended PBH mass function
as mentioned above, there are a number of issues that are not clear at the moment. For example, in the current modelling the motion
of PBHs are neglected, the correlations between monopoles and antimonopoles are ignored, etc. Also the study is restricted
to the case of a radiation-dominated universe and relatively high symmetry breaking scales ($\gtrsim\mathcal{O}(10^{10}\GeV)$).
These aspects are worth further explorations and being checked against numerical simulations which are needed to
determine the prospects of producing long-lived magnetic black holes and resolving the accompanying monopole problem.

\begin{acknowledgments}
Chen Zhang would like to thank Yi-Lei Tang for helpful discussion. The authors are grateful to the anonymous referee
whose report motivates an extended discussion about the difference between the two ways of modelling and related issues
which is presented in sections ~\ref{subsec:flux} and ~\ref{subsec:anatomy}. This work was supported by
the National Natural Science Foundation of China (Grants Nos. 11975072 and 11835009)
and the National SKA Program of China (Grants Nos. 2022SKA0110200 and 2022SKA0110203).
\end{acknowledgments}

\bibliography{mmpbh_v10}
\bibliographystyle{JHEP}

\end{document}